\documentclass{aa}  
\usepackage{graphicx}
\usepackage{txfonts}
\usepackage{upgreek}
\usepackage{tipa}
\usepackage{verbatim}
\usepackage{natbib}
\usepackage{color}
\newcommand{\uchii}{UC-H\scriptsize{II}\normalsize}
\newcommand{\micron}{$\upmu$m}
\newcommand{\lsol}{L$_\odot$\,}
\newcommand{\msol}{M$_\odot$\,}
\newcommand{\point}{\cdot}
\newcommand{\kms}{km$\cdot${s$^{-1}$}}
\newcommand{\kkms}{K$\cdot$km$\cdot${s$^{-1}$}}
\newcommand{\msolkmsy}{\msol$\cdot$km$\cdot${s$^{-1}$$\cdot$yr$^{-1}$}}
\newcommand{\mollinej}[2]{($J $={#1}$-${#2})}
\newcommand{\molline}[2]{({#1}$-${#2})}
\newcommand{\ttp}[1]{$\times 10^{#1}$}
\newcommand{\hcop}{HCO$^+$}
\newcommand{\ndhp}{N$_2$H$^+$}

\begin{document}
\newcommand{\cf}{\textit{see}}
   \title{Evolution of massive protostars: the IRAS~18151$-$1208 region\thanks{Observed with the IRAM 30m and Mopra telescopes. IRAM is supported by INSU/CNRS (France), MPG (Germany) and IGN (Spain). The Mopra telescope is part of the Australia Telescope which is funded by the Commonwealth of Australia for operation as a National Facility managed by CSIRO.}$^,$\thanks{Channel maps in FITS format of Fig.~2 are available in electronic form at the CDS via anonymous ftp to cdsarc.u-strasbg.fr (130.79.128.5) or via http://cdsweb.u-strasbg.fr/cgi-bin/qcat?J/A+A/xxx/xxx}}

   \subtitle{}

   \author{M. Marseille\inst{1,2} \and S. Bontemps\inst{1,2} \and F. Herpin\inst{1,2} \and F.~F.~S. van der Tak\inst{3} \and C.~R. Purcell\inst{4}}

   \offprints{M. Marseille}

   \institute{Universit\'e Bordeaux 1, Laboratoire d'Astrophysique de Bordeaux\\
   email: \texttt{[marseille;bontemps;herpin]@obs.u-bordeaux1.fr}
   \and
    CNRS/INSU, UMR 5804, BP 89, 33270 FLOIRAC, France.
   \and 
   SRON Netherlands Institute for Space Research, Landleven 12, 9747 AD Groningen, The Netherlands.\\
   email: \texttt{vdtak@sron.nl}
  \and
  University of Manchester, Jodrell Bank Observatory, Macclesfield, Cheshire SK11 9DL, United Kingdom\\
  email: \texttt{Cormac.Purcell@manchester.ac.uk}
   }

   \date{Received 18 March 2008, Accepted 3 June 2008}

 
  \abstract
{The study of physical  and chemical properties of massive protostars is critical to better understand the evolutionary sequence which leads to the formation of high-mass stars.}
{IRAS~18151$-$1208 is a nearby massive region ($d =3$~kpc, $L \sim 2$\ttp{4}~\lsol) which 
splits into three cores: MM1, MM2 and MM3 (separated by 1\arcmin--2\arcmin). We aim at (1) studying the physical and chemical properties of the individual MM1, MM2 and MM3 cores; 
(2) deriving their evolutionary stages; (3) using these results to improve our view of the evolutionary
sequence of massive cores.}
{The region was observed in the
CS, C$^{34}$S, H$_2$CO, HCO$^+$, H$^{13}$CO$^+$, and N$_2$H$^+$ lines at mm wavelengths with the IRAM
30m and Mopra telescopes. We use 1D and 2D modeling of the dust continuum to derive the density and temperature distributions, which are then used in the RATRAN code to model the lines and 
constrain the abundances of the observed species.}
{All the lines were detected in MM1 and MM2. MM3 shows weaker emission, or even is undetected in HCO$^+$ and all isotopic species. MM2 is driving a newly discovered CO outflow and hosts a mid-IR-quiet massive protostar. The abundance of CS is significantly larger in MM1 than in MM2, but smaller than in a reference massive protostar such as AFGL~2591. In contrast the N$_2$H$^+$ abundance decreases from MM2 to MM1, and is larger than in AFGL~2591.}
{Both MM1 and MM2 host an early phase massive protostar, but MM2 (and mid-IR-quiet sources in general) is younger and more dominated by the host protostar than MM1 (mid-IR-bright). The
MM3 core is probably in a pre-stellar phase. We find that the N$_2$H$^+$/C$^{34}$S ratio varies systematically with age in the massive protostars for which the data are available. It can be used to identify young massive protostars.

}

   \keywords{ISM: individual objects: IRAS~18151$-$1208 -- ISM: abundances -- Stars: formation -- Line: profiles}

   \maketitle
%

\section{Introduction}
	
How high-mass stars form is still an open issue
\citep[\textit{e.g.}][]{zinnecker2007}. It is particularly not clear whether the
formation process for OB/high-mass stars is different from the way
solar-type/low-mass stars form. Stars more massive than $\sim$10~\msol\ may form
like a scaled-up version (high accretion rates) of the single (or monolithic)
collapse observed for the low-mass stars, or require a more complex process in
which competitive accretion inside the central regions of a forming cluster may
play a decisive role \citep[\textit{e.g.}][]{bonnell2004}. In the first
scenario, the observed high-mass clumps (100 to 1000~M$_\odot$; 0.5~pc in size)
are expected to fragment to form self-gravitating high-mass cores (10 to
100~M$_\odot$; 0.01$-$0.1~pc in size; \textit{e.g.} \citealt{krumholz2007})
which would collapse individually to form a massive single or binary stars. In
the second scenario, the competitive accretion is expected to occur inside the
high-mass clumps. The study of the properties of high-mass clumps is therefore a
central observational issue to progress in our understanding of the earliest
phases of high-mass star formation.

From a selection of IRAS sources not associated with any bright radio source but
having the IRAS colors of \uchii\ regions \citep[as defined by][]{wood1989},
\citet{sridharan2002} have built a sample of so-called High Mass Protostellar
Objects (hereafter HMPOs) which would correspond to the pre-\uchii\ phase of the
formation of high-mass stars. \citet{beuther2002a} found that the HMPOs were
systematically associated with massive clumps as detected in the dust continuum
and in CS line emission. These clumps are good candidates to correspond to the
earliest phases of high-mass star formation. They have not yet formed any bright
H{\sc{ii}} regions and contain a large amount of gas at high densities. The
precise evolutionary stages of the individual HMPOs might however be very
diverse and they require to be derived individually through dedicated, detailed
studies.

Recently, \citet{motte2007} investigated the whole Cygnus~X complex and obtained
a first unbiased view of the evolutionary scheme for high-mass clumps and
cores. Like for the HMPOs, the massive clumps ($\sim$0.5~pc in size) in Cygnus~X
could be resolved into massive cores ($\sim$0.1~pc). Roughly half of these cores
were found to be very bright in the (mid)-infrared \citep[such as AFGL~2591;
\textit{e.g.}][]{vandertak1999}, and therefore very luminous: the mid-IR
high-luminosity massive cores. The other half are weak or not detected in the
mid-IR, hereafter the mid-IR-quiet massive cores. Surprisingly all the
mid-IR-quiet massive cores were however found to drive powerful SiO
outflows. They could therefore be safely understood as the precursors of the
infrared high-luminosity massive cores.

\begin{figure}[t!]
  	\centering
  	\includegraphics[width=\columnwidth]{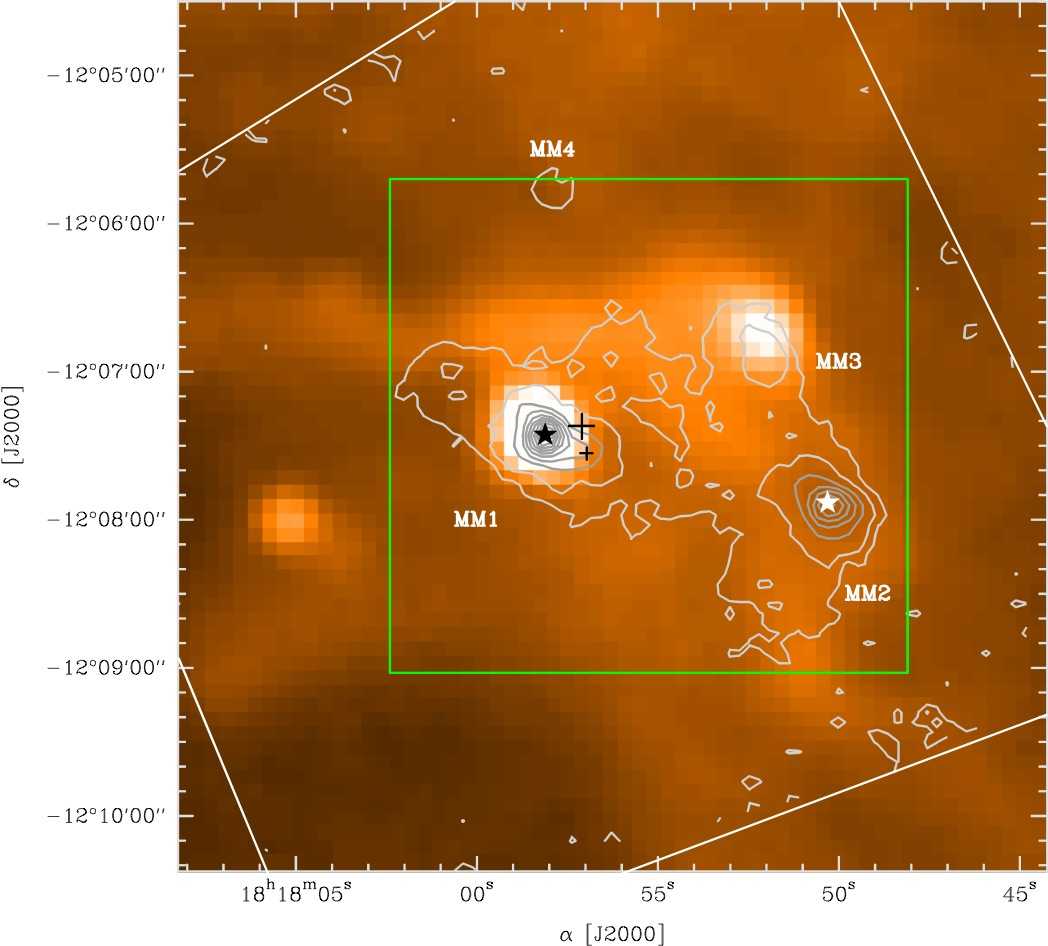}\\
      	\caption{Image of the MSX 8\micron\ emission toward IRAS~18151$-$1208 (color scale) overlaid with the 1.2~mm map (white polygon) by \citet{beuther2002a} (white contours at 5 and 10~\%, and grey contours from 20~\% to 90~\% of the maximum).
	The large and small crosses indicate the positions of the IRAS source and of MM1-SW (see text) respectively. 
	The black and white star symbols show the positions of the methanol and water masers respectively. The green polygon displays the region mapped in the present study (see Fig.~\ref{fig:maps}).}
         \label{fig:msx-map}
\end{figure}

IRAS~18151$-$1208 is a rather typical ($L \sim 10^4$~\lsol) and relatively
nearby (3~kpc) HMPO \citep{sridharan2002}. \citet{beuther2002a} show that the
clump actually splits into four individual cores MM1, MM2, MM3 and MM4 (see
Fig.~\ref{fig:msx-map} for an overview of the region). In addition MM1 seems to
further split into two cores, separated by 16\arcsec.  We hereafter refer to
MM1-SW for the secondary peak in the South-West of MM1 which possibly hosts an
embedded protostar \citep{davis2004}.  The weakest core MM4 is clearly outside
the main region, hence it will not be further considered in this paper.  A CO
outflow toward MM1 was discovered by \citet{beuther2002b}.  The MM1 and MM2
cores have sizes roughly two times larger, and similar masses (using the same
dust emissivity and temperature) than the most massive cores in
\citet{motte2007}. The IRAS source coincides with MM1 and no significant IRAS
contribution can be safely attributed to MM2 or MM3. MM3 is the least massive
and compact core and could actually be a still quiescent or pre-stellar
core. The IRAS~18151$-$1208 region is therefore particularly interesting to
study since it hosts three individual cores which could be interpreted as
high-mass star formation sites in three different evolutionary
stages. Millimeter and sub-millimeter wave observations are reported in
\citet{mccutcheon1995} and \citet{beuther2002a}. The near-IR counterparts (H$_2$
jets and HH objects) of the CO outflow driven by MM1 have been imaged by
\cite{davis2004}.

Molecular line observations compared with results of line modeling based on a physical model of the source is a classical 
technique to constrain physical and chemical properties of protostellar objects \citep[see][for a review]{ceccarelli1996,vandertak2005}.
While different approaches can be adopted, the most reliable and most often used method consists in constraining first 
the physical (density and temperature distributions) model from the dust continuum
emission (mid-IR to millimeter wavelengths), and then use this physical model to derive the fractional abundances of molecular species
from line emission modeling \citep{vandertak1999, vandertak2000, schoier2002, belloche2002, hatchell2003}.
Due to the dramatic changes in physical conditions inside the protostellar envelopes (increase of density and temperature, radiation, shocks) chemical evolution is observed and can be modeled thanks to chemical network codes. It is even expected that chemistry could 
provide a reliable clock to date protostellar objects \citep[\textit{e.g.}][]{vandishoeck1998,doty2002,wakelam2004}.
			
After the description of the molecular observations and of the continuum data from the literature (Sect.~2), the results are presented in Sect.~3. Section~4 details the modeling procedure from the fits of the spectral energy distributions (hereafter SEDs) for MM1 and MM2 using MC3D\footnote{See \citet{wolf1999} for details.} code (Sect.~4.1.1 and 4.2.1) to the non-LTE calculations of the line profiles and intensities for all observed molecules using RATRAN\footnote{See \citet{vandertak1999,vandertak2000b} for details.} code (Sect.~4.1.2, 4.1.3, 4.2.2, and 4.2.3). In Section~5, we discuss the results of this detailed analysis in order to improve our observational view of the evolution of high-mass cores from pre-stellar to high luminosity massive protostars.

\begin{table}[t!]
\caption{IRAS~18151$-$1208 sources characteristics. Offsets from IRAS position, J2000 coordinates and velocity in the local standard of rest \citep{beuther2002a} are reported.}             
\label{tab:sources}      
\begin{tabular}{lccccc}
\hline
\hline
source & $\Delta\alpha$[\arcsec]& $\Delta\delta$[\arcsec] & $\alpha$[J2000] & $\delta$[J2000]  & $\varv$ [km/s] \\
\hline 
MM1 & 13.2 & -4.9 & 18$^h$17$^m$58.0$^s$ & -12$^\circ$07'27" & 33.4  \\
MM2 & -98.9 & -32.8 & 18$^h$17$^m$50.4$^s$ & -12$^\circ$07'55" & 29.7  \\
MM3 & -72.3 & 26.5 & 18$^h$17$^m$52.2$^s$ & -12$^\circ$06'56" & 30.7  \\
\hline
\end{tabular}
\end{table}


\section{Observations.}

\begin{table*}[t!]
\begin{minipage}[t]{500pt}
\caption{List of observational parameters for IRAM 30m and 22m Mopra telescopes. Here are indicated observed species, energy level transitions, line emission frequencies, half power beam width (HPBW), instrument (30m for IRAM-30m and Mo for Mopra telescope), main beam efficiency $\eta_{mb}$, receiver name, velocity resolution $\delta \varv$, observation mode and system temperature $T_{sys}$.}           
\label{table4}      
\renewcommand{\footnoterule}{}
\begin{center}
\begin{tabular}{lccccccccc}
\hline
\hline
Species 	& Transition	& Frequency	& HPBW	& Instrument$^d$	& $\eta_{mb}$ & Receiver	& $\delta \varv$     & Mode & $T_{sys}$\\ 
		&		 	& [GHz] 		& [\arcsec]		& &		         &                          & [m$\point$s$^{-1}$]  &            &  [K]             \\
\hline
CS				& $J=2-1$ & 97.9809 & 25 & 30m & 0.78 & A100 & 60 & RaM$^a$  & 260 \\
				& $J=2-1$ & 97.9809 & 32 & Mo & 0.49 & 3.5mm & 191 & OTF$^c$ & 240  \\
				& $J=3-2$ & 146.9690 & 17 & 30m & 0.69 & D150 & 80 & RaM$^a$  & 420 \\
				& $J=5-4$ & 244.9356 & 10 & 30m & 0.49 & D270 & 96 & RaM$^a$  & 1150 \\
\hline
C$^{34}$S			& $J=2-1$ & 96.4129 & 25 & 30m & 0.78 & A100 & 61 & RaM$^a$  & 190 \\
				& $J=2-1$ & 96.4129 & 32 & Mo & 0.49 & 3.5mm & 194 & OTF$^c$ & $\sim$270 \\
				& $J=3-2$ & 144.6171 & 17 & 30m & 0.69 & D150 & 81 & RaM$^a$  & 390 \\
\hline
CO				& $J=2-1$ & 230.5380 & 11 & 30m & 0.52 & HERA & 406 & RaM2$^b$  &$\sim$600\\
\hline
H$_2$CO 			& $J=3_{03}-2_{02}$ & 218.2222 & 11 & 30m & 0.54 & HERA & 107 & RaM2$^b$  & 250 \\
				& $J=3_{22}-2_{21}$ & 218.4756 & 11 & 30m & 0.54 & HERA & 107 & RaM2$^b$  & 250 \\
				& $J=3_{12}-2_{11}$ & 225.6978 & 11 & 30m & 0.53 & A230 & 104 & RaM$^a$  & 870 \\
\hline
HCO$^+$ 	& $J=1-0$ & 89.1885 & 36 & Mo & 0.49 & 3.5mm & 210 & OTF$^c$ &  $\sim$260\\
H$^{13}$CO$^+$ 	& $J=1-0$ & 86.7542 & 36& Mo & 0.49 & 3.5mm & 216 & OTF$^c$ & $\sim$175\\
\hline
N$_2$H$^+$ & $J=1-0$ & 93.1737 & 36 & Mo & 0.49 & 3.5mm & 201 & OTF$^c$ &  $\sim$180\\
\hline
& & & & &  \\
\end{tabular}\\
\end{center}
$^a$ RaM -- Raster Map mode set to create a 3~$\times$~3 pixels mini-map with a $\Delta\theta$ step of 10\arcsec\ around source emission peak.\\
$^b$ RaM2 -- Raster Map mode with HERA receiver, set to create a 190\arcsec~$\times$~140\arcsec\ map with a $\Delta\theta$ step of 6\arcsec\ towards MM1, MM2 and MM3.\\
$^c$ OTF -- On-The-Fly observing mode carrying out on a 5\arcmin~$\times$~5\arcmin\ grid with an angle step $\Delta\theta$ of 9\arcsec\ and an OFF position at 30\arcmin\ from the center.\\
$^d$ Conversional factor is $S/T_{mb}=4.95$~Jy/K for IRAM 30m telescope and $S/T_{mb}=22$~Jy/K for ATNF 22m telescope.
\end{minipage}
\end{table*}

Observations were performed during two sessions. The first one in June, 2005 at Mopra, the 22m dish of the Australia Telescope National Facility\footnote{http://www.narrabri.atnf.csiro.au/mopra/} (ATNF). The second one in August, 2005 at Pico Veleta, on the IRAM 30m antenna\footnote{http://iram.fr/IRAMES/index.htm}.

\subsection{Mopra observations}

	The IRAS~18151$-$1208 region was mapped in the On-The-Fly (OTF) mode in the rotational transition of CS\,\mollinej{2}{1}\ at 98.0~GHz, C$^{34}$S\,\mollinej{2}{1}\ at 96.4~GHz , HCO$^+$\,\mollinej{1}{0}\ at 89.2~GHz, H$^{13}$CO$^+$\, \mollinej{1}{0}\ at 86.8~GHz and N$_2$H$^+$\,\mollinej{1}{0} at 93.2~GHz.  Each run covered a field of 5\arcmin~$\times$~5\arcmin\ with a half-power beam width (HPBW) of approximatively 36\arcsec\ ($\eta_{mb}=0.49$) and sampled with a $\Delta \theta$ step of 9\arcsec. We used the 3~mm receiver of the ATNF telescope with the digital auto-correlator set to the 64~MHz bandwidth with 1024 channels and both polarizations. It provides a velocity resolution of $\sim$0.2~\kms\ over a 200~km$\point$s$^{-1}$ range. Observations happened while weather was clear, with a system temperature $T_{sys}$~$\sim$~175 -- 260~K (see Table~\ref{table4} for details). Pointing and focus adjustments have been set on Jupiter or known sources of calibration.

	Raw data have been reduced with AIPS++ LIVEDATA and GRIDZILLA tasks \citep{mcmullin2004}. LIVEDATA performs the subtraction between the SCAN spectra row and the OFF spectrum. Then it fits the baseline with a polynomial. We used the GRIDZILLA package to resample and build a data cube with regular pixel scale, weighted with the system temperature $T_{sys}$. We created a 39~$\times$~39 grid of 9\arcsec~$\times$~9\arcsec\ pixel size convolved by a Gaussian with a FWHM of 18\arcsec, truncated at an angular offset of 36\arcsec. Finally, due to positional errors that we found in maps, a shift in $\alpha$ of 14\arcsec\ for \ndhp, 10\arcsec\ for CS and 5\arcsec\ for HCO$^+$ has been applied to fit millimeter-continuum peak positions of the three sources given by \citet{beuther2002a}.

\subsection{IRAM-30m observations}

	We observed the rotational transition of CO\,\mollinej{2}{1} molecule at 230.5~GHz and the two H$_2$CO rotational para-transitions \mollinej{$3_{03}$}{$2_{02}$} and \mollinej{$3_{22}$}{$2_{21}$} at respectively 218.2~GHz and 218.5~GHz simultaneously, using the HERA 3~$\times$~3 pixel dual multi-beam receiver with VESPA backends at a resolution of 160~kHz for CO and 40~kHz for H$_2$CO. Thus we built two 190\arcsec~$\times$~140\arcsec\ maps sampling every 6\arcsec\ around MM1, MM2 and MM3. This observation was done while weather was good ($\tau_0^{atm}=0.17$), with $T_{sys}\sim 600$~K and 250~K for CO and H$_2$CO respectively. The maps are incomplete because raster mapping was not set to cover the entire $\Delta\alpha$ and $\Delta\delta$, and the HERA vertical polarization (set for the CO) had a dead pixel (number 4). All pointings and focus have been set on suitable planets (\textit{i.e.} Jupiter) or calibration sources.

	The IRAS~18151$-$1208 region was also mapped in the rotational transitions of CS at 98.0, 147.0 and 244.9~GHz ($J$=2$-$1, $J$=3$-$2, $J$=5$-$4), in the isotopic C$^{34}$S rotational transitions at 96.4 and 144.6~GHz ($J$=2$-$1, $J$=3$-$2)  and in the H$_2$CO rotational ortho-transition ($J$=$3_{12}-2_{11}$) at 225.7~GHz, using simultaneously A100, A230, D150  and D270 receivers coupled to the high-resolution VESPA backend (resolution of 80~kHz).  Central emission peaks of MM1, MM2 and MM3 were quickly mapped with a 3~$\times$~3 raster position switching method with a step of 10\arcsec. Observations happened while sky conditions were reasonable ($\tau_{0}^{atm}\sim 0.1$--0.6 and T$_{sys}\sim 260$--1150~K).
	
	 Reduction for all IRAM 30m telescope data were performed with the CLASS software from the GILDAS suite \citep{guilloteau2000}. We found and eliminated unusable data and treated platforming that appeared in CO and H$_2$CO baselines, then spectra at the same position were summed and finally antenna temperature $T_a^*$ was converted into $T_{mb}$ (see Table~\ref{table4} for the values of $\eta_{mb}$).


\section{Results}

The modeling of the spectral energy distribution (SED) of a source is a common way to derive its density and temperature profiles \citep[\textit{e.g.}][]{vandertak1999}. 
The obtained so-called physical model is then used as such to derive the abundances and to further investigate the physical properties 
(kinematics for instance) through the line emission modeling. 

We thus observed CS, C$^{34}$S, HCO$^+$ and H$^{13}$CO$^+$ line emissions which are bright and are tracing dense gas. We also observed H$_2$CO line emission which traces dense gas and its temperature. In additions we observed N$_2$H$^+$ line emission which is a cold gas indicator and finally CO that traces molecular gas flows.


\subsection{Large scale maps: CO\,\molline{2}{1}, H$_2$CO\,\molline{$3_{22}$}{$2_{21}$}, CS\,\molline{2}{1}, HCO$^+$\,\molline{1}{0}, H$^{13}$CO$^+$\,\molline{1}{0} and N$_2$H$^+$\,\molline{1}{0}}

\begin{figure*}[t!]
   \includegraphics[width=500pt]{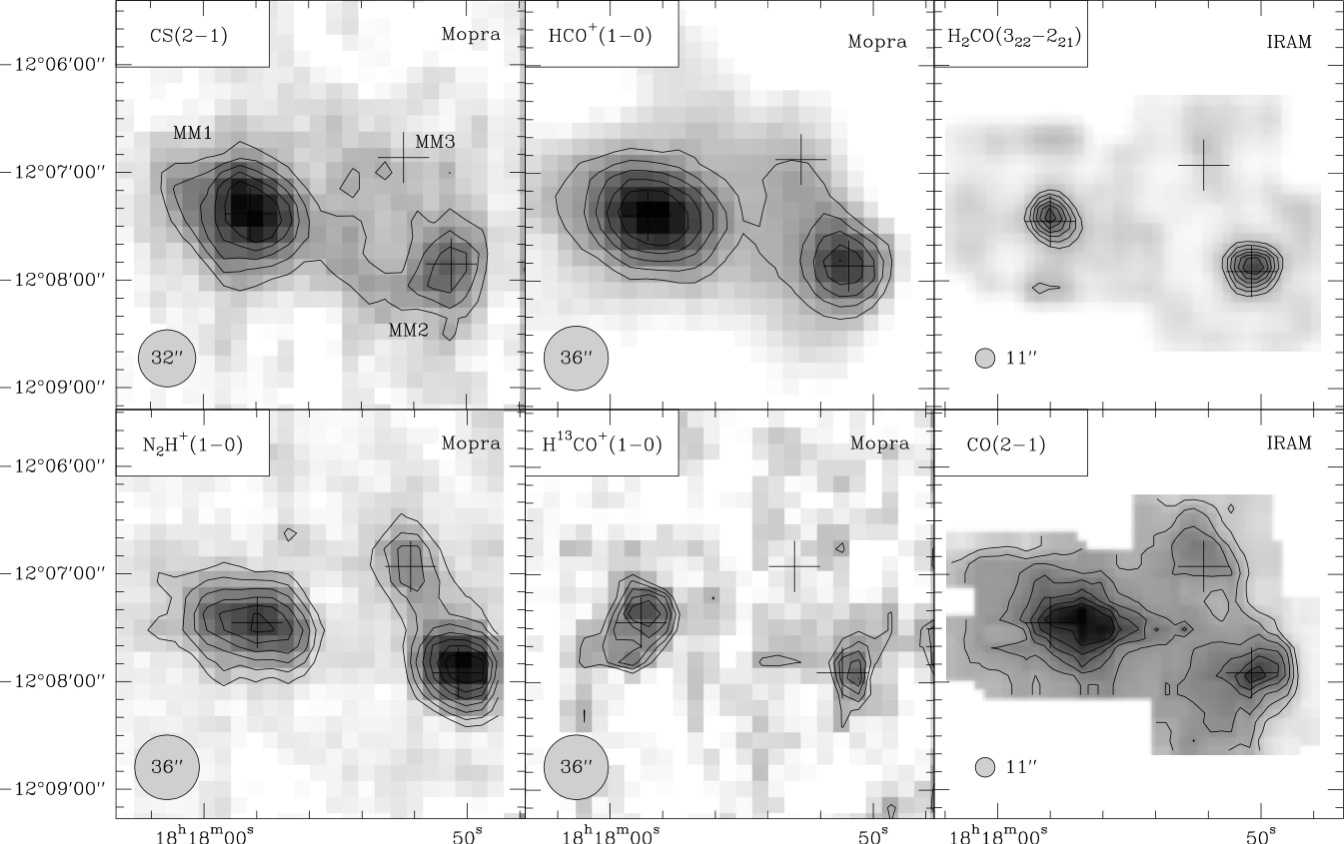}
      \caption{Integrated velocity maps of CS\,\molline{2}{1}, N$_2$H$^+$\,\molline{1}{0}, HCO$^+$\,\molline{1}{0}, H$^{13}$CO$^+$\,\molline{1}{0} obtained with the Mopra 22m telescope (left and middle panels) and CO\,\molline{2}{1}, H$_2$CO\,\molline{$3_{22}$}{$2_{21}$} maps obtained with the IRAM 30m telescope (right panel). Level contours fit 90~\% to 10~\% of peak flux with a 10~\% step. Only levels exceeding the threshold of 2$\sigma_{noise}$ in each map are plotted in order to identify detections.  Peak fluxes are 16.9~K$\point$km$\point$s$^{-1}$ for CS,  16.9~K$\point$km$\point$s$^{-1}$ for N$_2$H$^+$, 32.9~K$\point$km$\point$s$^{-1}$ for HCO$^+$, 2.31~K$\point$km$\point$s$^{-1}$ for H$^{13}$CO$^+$, 21.1~K$\point$km$\point$s$^{-1}$ for H$_2$CO and 393~K$\point$km$\point$s$^{-1}$ for CO. Crosses indicate the 1.2~mm continuum positions \citep{beuther2002a} for the three sources. 
      }
         \label{fig:maps}
   \end{figure*}

The velocity integrated maps in CO\,\molline{2}{1}, H$_2$CO\,\molline{$3_{22}$}{$2_{21}$}, CS\,\molline{2}{1}, HCO$^+$\,\molline{1}{0},
H$^{13}$CO$^+$\,\molline{1}{0} and N$_2$H$^+$\,\molline{1}{0} are displayed in Fig.~\ref{fig:maps}.
Table~\ref{tab:extension} gives the derived emission extensions (at the 3$\sigma_{\mathrm{noise}}$ level), 
the minor-to-major axis ratio $b/a$ and
position angle (P.A.) of MM1, MM2 and MM3. Note that the C$^{34}$S\,\molline{2}{1} MOPRA map was too noisy and was discarded.

\begin{table*}[t!]
\caption{Sources extension at 3$\sigma_{\mathrm{noise}}$, minor-to-major axis ratio $b/a$ and position angle (P.A.; counterclockwise angle relative to the North) 
for each molecular transition observed. When the emission is almost spherical ($b/a \geq 0.85$) P.A. is uncertain and set to 0$^\circ$.}
\begin{center}
\begin{tabular}{@{\extracolsep{10pt}}l ccc ccc ccc}
\hline
\hline
				& \multicolumn{3}{c}{MM1}				& \multicolumn{3}{c}{MM2}				& \multicolumn{3}{c}{MM3}\\
\cline{2-4}\cline{5-7}\cline{8-10}
				& Ext.				&	$b/a$ & P.A. [$^\circ$]		& Ext.		&	$b/a$ &  P.A. [$^\circ$]		& Ext.		&	$b/a$ & P.A. [$^\circ$] \\
\hline
CO\,\molline{2}{1}	& 72\arcsec~$\times$~51\arcsec	& 0.71	& +85 	& 34\arcsec~$\times$~23\arcsec	& 0.68	& -53	& 45\arcsec~$\times$~36\arcsec	& 0.8	& +53 \\
H$_2$CO\,\molline{$3_{22}$}{$2_{21}$} & 35\arcsec~$\times$~30\arcsec	&	0.85 & 0	& 35\arcsec~$\times$~30\arcsec &	0.85 & 0	& --		& -- & -- \\

CS\,\molline{2}{1}	& 97\arcsec~$\times$~81\arcsec	&	0.83 & +57		& 75\arcsec~$\times$~62\arcsec	&	0.83 & -31	& --		& -- & -- \\
N$_2$H$^+$\,\molline{1}{0}	& 89\arcsec~$\times$~62\arcsec	&	0.69 & +77	& 55\arcsec~$\times$~54\arcsec	&	0.98 & 0	& 54\arcsec~$\times$~52\arcsec & 0.96 & 0 \\
HCO$^+$\,\molline{1}{0}	& 104\arcsec~$\times$~78\arcsec	&	0.75	& +75 	& 57\arcsec~$\times$~56\arcsec	&	0.98 & 0 & --		& -- & -- \\
H$^{13}$CO$^+$\,\molline{1}{0}	& 52\arcsec~$\times$~38\arcsec	&	0.73 & -28	&  $< HPBW$	&	-- & --	& --		& -- & -- \\

\hline
\end{tabular}
\end{center}
\label{tab:extension}
\end{table*}

The molecular emission always peaks at the positions of the three cores (millimeter-continuum peak positions by \citealt{beuther2002a}). 
While MM1 and MM2 are always well detected, MM3 seems to be too weak in CS\,\molline{2}{1}, H$_2$CO\,\molline{$3_{22}$}{$2_{21}$} and 
H$^{13}$CO$^+$\,\molline{1}{0} to be detected. The emission is extended even between the cores for CS, HCO$^+$ and CO lines but are more
peaked for N$_2$H$^+$, H$^{13}$CO$^+$ and H$_2$CO lines. Some chemical differences are already clear in CS, N$_2$H$^+$ and H$_2$CO. While MM1 is generally brighter than MM2, such as in the dust continuum, MM2 has the same intensity in H$_2$CO and is even significantly brighter in N$_2$H$^+$ than MM1.

CS\,\molline{2}{1} line emission is detected for MM1 and MM2 and weak emission could be present at 1$\sigma$ for MM3. MM1 emission is stronger than MM2, and covers a larger area, with extensions of 97\arcsec~$\times$~81\arcsec\ and 75\arcsec~$\times$~62\arcsec\ for MM1 and MM2 respectively. It indicates that a
large scale envelope is surrounding MM1 and MM2. Axis ratios $b/a$ are equal to 0.83 for MM1 and MM2, showing that the emission is almost
spherical. The position angle of MM1 emission ($+57^{\circ}$) is similar to the position angle of the large dust emission of $+62^{\circ}$ given by
\citet{beuther2002a}. We note that emission is detected between MM1 and MM2. It suggests that they could be connected by a dense molecular filament.

The cold gas tracer N$_2$H$^+$\,\molline{1}{0} is detected toward the three cores. Emission from MM2 is stronger and more concentrated than for MM1
(89\arcsec~$\times$~62\arcsec\ and 55\arcsec~$\times$~54\arcsec\ respectively), suggesting that MM2 has a smaller and colder envelope. A similar behaviour
has been observed by \citet{reid2007} for two similar cores in IRAS~23033$+$5951. The elongation of MM1 ($b/a = 0.69$) and position angle ($+77^{\circ}$) ressembles the extension of the dust emission. MM3 line emission is stronger
compared to other molecules and shows a connection with MM2. Together with a similar rest velocity for MM2 and MM3, it confirms that MM3 belongs to
the IRAS~18151$-$1208 region.

HCO$^+$\,\molline{1}{0} and H$^{13}$CO$^+$\,\molline{1}{0} lines are detected for both MM1 and MM2. MM3 has a weak emission in
HCO$^+$\,\molline{1}{0}. As for CS, MM1 emission is stronger and more extended than for MM2 (respective extensions are 104\arcsec~$\times$~78\arcsec\
and 57\arcsec~$\times$~56\arcsec). With $b/a=0.75$ and a position angle of $+75^{\circ}$, we note that MM1 emission may be extended along the outflow
as expected since the HCO$^+$ line emission is definitely showing outflow wings such as in CO. The map shows an emission between MM1 to MM2, and MM2 to MM3, confirming that the three sources are connected.

Dense gas traced by H$_2$CO\,\molline{$3_{22}$}{$2_{21}$} line is detected for MM1 and MM2 only. The emission from each source coincides perfectly with
the dust continuum positions. MM1 and MM2 have similar intensities and have roughly the same small circular size (35\arcsec~$\times$~30\arcsec,
$b/a=0.85$ for the two sources) and do not show any spatial suggestions of being contaminated by outflows. 

The CO\,\molline{2}{1} line is detected in all sources and shows an extended emission from the whole region. MM1 emission is extended to the west
(72\arcsec~$\times$~51\arcsec, $b/a = 0.71$ and P.A.$= +85^{\circ}$) due to the red wing of the outflow (P.A.$= +88^{\circ}$). See the following section for more details on the CO outflows.

\subsection{Molecular outflows in CO\,\molline{2}{1}}

\begin{figure}[t!]
   \centering
   \includegraphics[width=200pt]{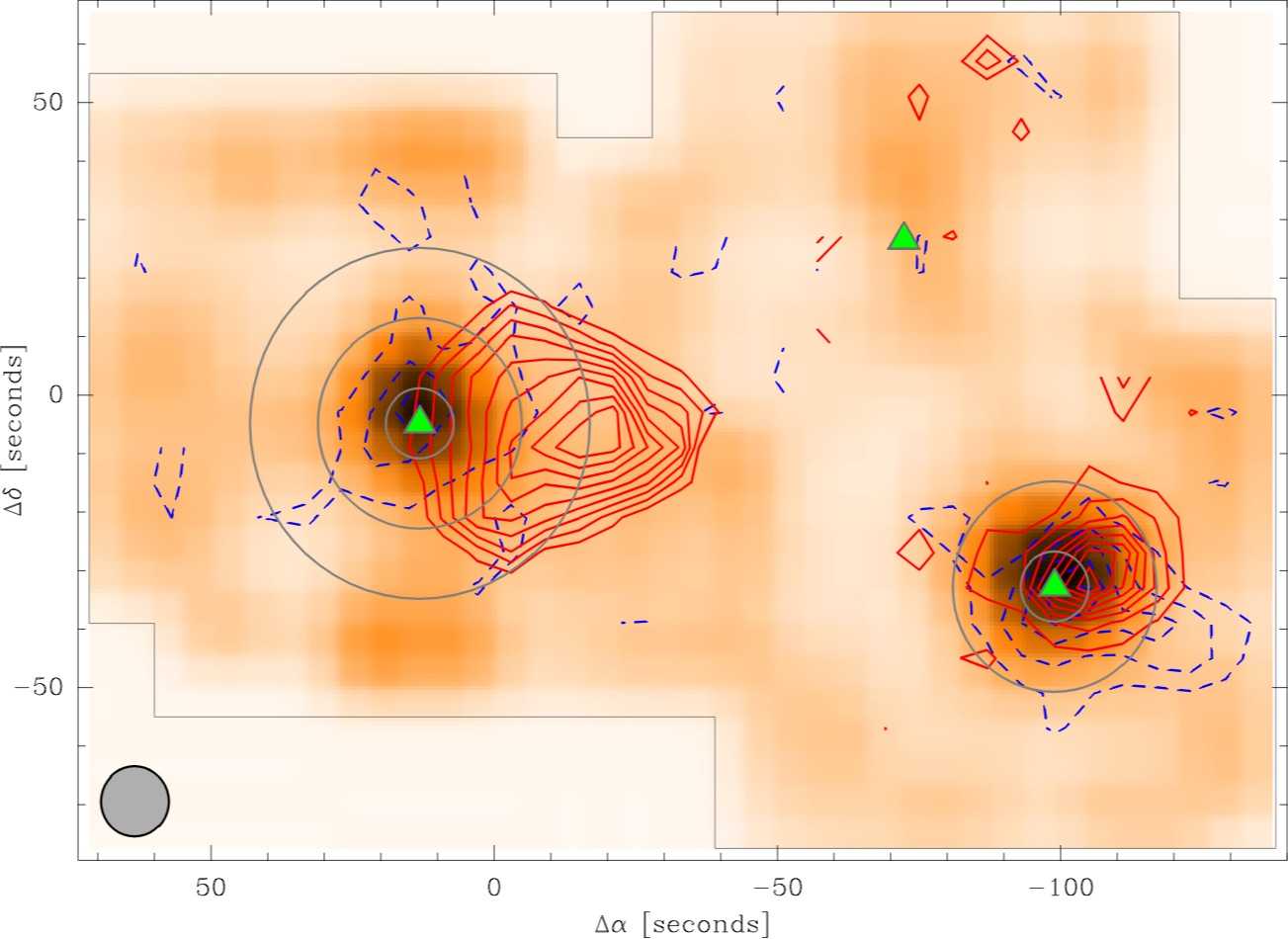}\\
   \includegraphics[width=200pt]{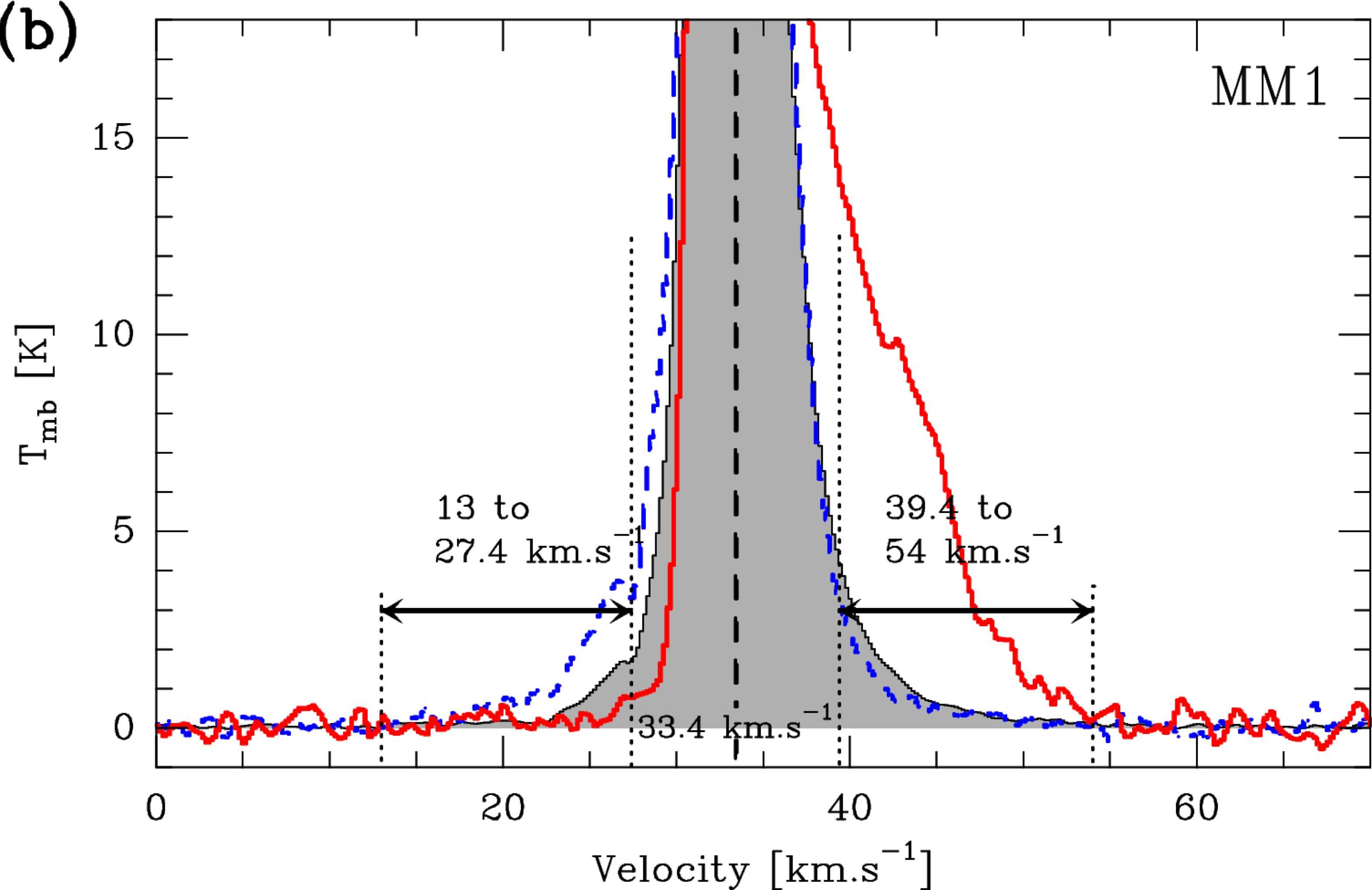}\\
   \includegraphics[width=200pt]{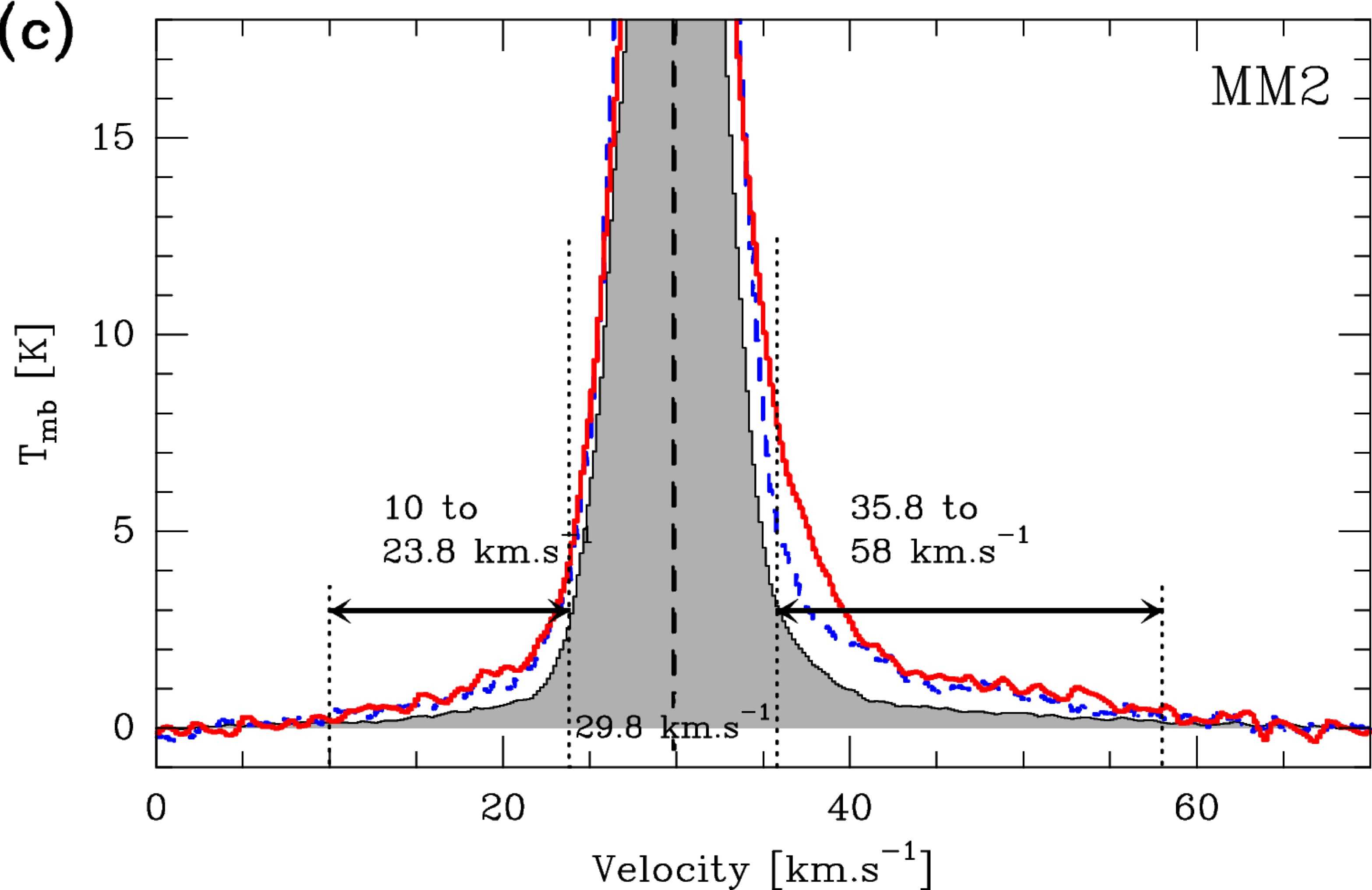}
   \caption{(a) Contour maps of the CO\,\molline{2}{1} blue-shifted (dashed) and
     red-shifted (solid) integrated emission in the outflow wings for MM1, MM2,
     and MM3 (no detection). Positions of the MM cores are marked by filled
     (green) triangles. Blue- and red-shifted velocity intervals are +13.0 to
     +27.4 and +39.4 to +54.0~\kms\ for MM1, +10.0 to +23.8 and +35.8 to
     +58.0~\kms\ for MM2 and MM3. Lowest, highest contours and contour steps are
     (10, 70, 15)~\kkms\ (blue) and (60, 410, 50)~\kkms\ (red) for MM1, and (10,
     70, 15)~\kkms\ (blue) and (20, 195, 25)~\kkms\ (red) for MM2 and MM3. The
     background grey-scaled image is the integrated H$_2$CO emission from
     Fig.~\ref{fig:maps}. (b) CO\,\molline{2}{1} spectra for the MM1 outflow
     together with the velocity cuts used to integrate maps (dotted vertical
     lines) in (a) and to derive the mechanical forces. The solid line circles
     around MM1 and MM2 are 6\arcsec and 12\arcsec\ radius and indicate the area
     on which the $F_{\mathrm{CO}}$ measurements are derived (see text for
     details). The filled grey spectrum is an average over the region of bright
     intensities in the wings as seen in the integrated map in (a). The dashed
     and blue (respectively solid and red) spectrum shows the line profile at
     the maximum of the blue (resp. red) lobe. (c) same as (b) for MM2.}
         \label{velocitymaps}
   \end{figure}

The CO\,\molline{2}{1} line emission shows clear outflow wings. They are due to
the already known CO outflow driven by MM1 \citep{beuther2002b} but seem also to
be located in the region of MM2. Figure~\ref{velocitymaps} displays the average
CO\,\molline{2}{1} spectra around MM1 and MM2, as well as the contour map of the
integrated emission in the wings. The displayed spectra clearly show emission
wings up to high velocities. The observed maximum velocities are -20.4 and
+20.6~\kms\ for MM1 and -19.7 and +28.3~\kms\ for MM2, with respect to
$\varv_{LSR}$ (see Fig.~\ref{velocitymaps} and Table~\ref{tab:outflow}). It is
clear from the map of this wing emission that there are actually two outflows
driven by the two respective cores. The MM2 outflow is actually newly discovered
(see Sect.~5.1 for a discussion).

The MM1 outflow appears mostly east-west oriented. This is not quite the
orientation observed for the main jet-outflow system recently discussed by
\citet{davis2004} with a NW-SE direction for the H$_2$ jet (see their Fig.~2). A
second H$_2$ jet, west of MM1 and probably driven by an embedded source in
MM1-SW (see Fig.~\ref{fig:msx-map}), is however also recognised in this
work. The position of MM1-SW is marked in Fig.~\ref{velocitymaps} with a small
green triangle. Its location is close to a possible secondary center of outflow
with a weak blue-shifted lobe in the south and the bright red-shifted western
lobe being curved toward that location. The spatial resolution of our CO
observation is not high enough to fully conclude about the association of CO
outflowing gas with the different driving sources, but it seems that the western
part of the MM1 outflow might actually be dominated by a second outflow from
MM1-SW. The inclination of the MM1 flow on the sky is difficult to
determine. From the significant extension of the flow and from the overlap of
the lobes we can only say that the flow is neither face-on (in the plane of the
sky) nor pole-on, and is therefore mostly intermediate.

The MM2 outflow is more compact than for MM1. Both lobes are confined in the
core. The integrated emissions in the two lobes are similar in strength and are
four times larger than the blue lobe of MM1. If the flow is dominated by a
single ejection, the large overlap of the two lobes suggests that the flow is
not face-on, but could be close to pole-on.

All the derived properties of the two outflows are summarized in
Table~\ref{tab:outflow}. In addition, to quantify the energetics of the two
detected outflows, we derive and give in this table the outflow forces using the
procedure described in ¥\citet{bontemps1996}. The mechanical force (or momentum
flux) seems to be the measurable quantity from observed CO outflows which can
approach in a not too uncertain way the corresponding quantity of the inner jet
or wind (\citealt{cabrit92}; \citealt{chernin95}). 
We estimate the outflow force $F_{\mathrm{CO}}$ for MM1 and MM2 by integrating
the momentum flux inside volumes with radius 6$''$ and 18$''$ centered on the
sources.
In addition, for the extended red lobe of MM1, a measurement is given between
18\arcsec\ and 30\arcsec\ to better cover the red lobe. Since the inclination of
the two outflows can not be reliably derived from such low resolution
observations we adopt the average correction of a factor of 10 adopted by
\citet{bontemps1996}.

 \begin{table*}[t!]
\begin{center}
\begin{tabular}{cccccccc}
\hline
\hline
 & $\Delta \varv_b$ & $\Delta \varv_r$ & $F_{\mathrm{CO_{b}}}$ & $F_{\mathrm{CO_{r}}}$ & $F_{\mathrm{CO_{total}}}$ & Ext. blue & Ext. red \\
 & [\kms] & [\kms] & [\msolkmsy] & [\msolkmsy] & [\msolkmsy] & & \\
\hline
MM1 & (13.0,27.4) & (39.4,54.0) & 0.44~$\pm$~0.12~\ttp{-3} & 1.86~$\pm$~0.11~\ttp{-3} & 2.30~$\pm$~0.23~\ttp{-4} & 17\arcsec~$\times$~11\arcsec  & 46\arcsec~$\times$~45\arcsec \\
MM2 & (10.0,23.8) & (35.8,58.0) & 0.48~$\pm$~0.09~\ttp{-3} & 1.86~$\pm$~0.15~\ttp{-3} & 2.34~$\pm$~0.24~\ttp{-4}  & 40\arcsec~$\times$~25\arcsec & 26\arcsec~$\times$~22\arcsec \\
\hline
\end{tabular}
\end{center}
\caption{Momentum flux of CO\,\molline{2}{1} calculated as in \citet{bontemps1996}. Blue and red integration intervals $\Delta \varv_{b,r}$ (\kms), blue-shifted momentum flux between 6\arcsec\ and 12\arcsec\, $F_{\mathrm{CO_{b}}}$ (\msolkmsy),  red-shifted momentum flux between 6\arcsec\ and 12\arcsec\, $F_{\mathrm{CO_{r}}}$ (\msolkmsy), total momentum flux $F_{\mathrm{CO_{total}}}$ (\msolkmsy), maximal and outflows extensions are reported for MM1 and MM2.
}
\label{tab:outflow}
\end{table*}
   
 
\subsection{Raster maps in CS, C$^{34}$S, and H$_2$CO\,\molline{$3_{12}$}{$2_{11}$}}

  \begin{figure}[t!]
   \centering
   \includegraphics[width=\columnwidth]{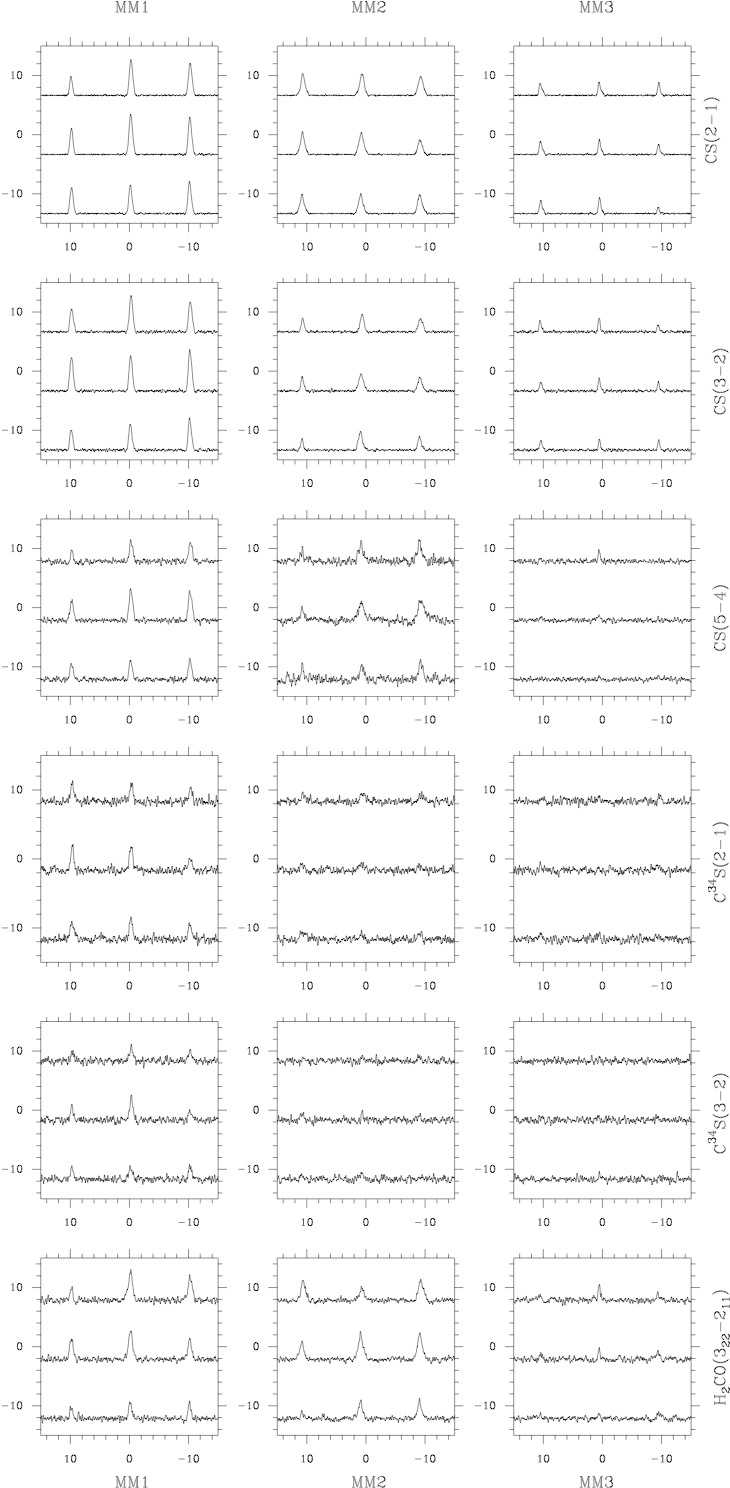}
   \caption{3~$\times$~3 maps in CS, C$^{34}$S, and
     H$_2$CO\,\molline{$3_{12}$}{$2_{11}$} for MM1, MM2 and MM3
     sources. Velocity ranges from $15$ to $50$~\kms\ in all spectra, intensity
     ranges from $-2$ to $5$~K for H$_2$CO line emission, $-1$ to $2$~K for
     C$^{34}$S line emissions, $-2$ to $5$~K for CS\,\molline{5}{4} line
     emission and $-2$ to $10$~K for other CS line emission. Spectra have been
     smoothed to an identical velocity resolution of 0.2~\kms.}
         \label{fig:raster}
   \end{figure}
 
   In addition to the large maps, a complementary investigation of the three
   millimeter sources is provided by the 3~$\times$~3 raster maps of CS,
   C$^{34}$S and H$_2$CO (see Fig.~\ref{fig:raster}).

All transitions are detected toward MM1 at $\varv_{LSR}=33.4\pm 0.1$~\kms. The
line profiles are similar with a typical line width (FHWM) $\Delta \varv=2.7\pm
0.2$~\kms. CS\,\molline{2}{1}, CS\,\molline{3}{2} and CS\,\molline{5}{4} have
roughly equal intensities. The optically thin C$^{34}$S\,\molline{2}{1} line
emission is detected all around the source at an approximately constant
intensity level of $\sim 1$~K. In contrast C$^{34}$S\,\molline{3}{2} seems to
peak on the central position. The line profiles and the spatial patterns for
H$_2$CO\,\molline{$3_{12}$}{$2_{11}$} and CS\,\molline{5}{4} are very similar,
suggestive of common origin and excitation conditions for the two transitions
($E_{\textrm{up}}/k$ are respectively equal to 33.40 and 35.27~K).

The intensities of most transitions toward MM2 are weaker than MM1. The line
profiles at $\varv_{LSR}=29.8 \pm 0.1$~\kms~have a triangular shape and a larger
line width ($\Delta \varv = 3.7\pm 0.1$~\kms) compared to MM1. This could be due
to a gas motion contribution through its turbulence. The intensity of
CS\,\molline{5}{4} is low compared to MM1, contrary to
H$_2$CO\,\molline{$3_{12}$}{$2_{11}$} line emission which is as bright as in
MM1.

All molecular line profiles are weaker in MM3, and seem to be less peaked than
in MM1 and MM2 as seen in velocity integrated maps
(cf. Fig.~\ref{fig:maps}). They are detected at all positions in
CS\,\molline{2}{1}, (3$-$2), they have a weak emission in two positions in
H$_2$CO\,\molline{$3_{12}$}{$2_{11}$} and finally CS\,\molline{5}{4} and the
C$^{34}$S transitions could not be detected.


\subsection{Continuum fluxes}
\label{ssec:sed}

\begin{table*}[t!]
\caption{Continuum flux densities of the three cores within IRAS~18151$-$1208 region. MSX flux densities are extracted from the catalog for MM1 and a 3$\sigma_{noise}$ calculation over the beam is done to derive an upper limit for the undetected MM2 and MM3 sources. IRAS measurements are derived from maps improved by HiRes method. Millimeter and sub-millimeter continuum flux densities are adapted to source reference size derived from continuum map by \citet{beuther2002a}.}             
\label{table2}      
\begin{center}       
\begin{tabular}{@{\extracolsep{1pt}}ccccccccccccccc}
\hline
\hline
~	& \multicolumn{4}{c}{MSX flux densities [Jy]}	& \multicolumn{4}{c}{IRAS flux densities [Jy]}	& \multicolumn{4}{c}{mm-continuum flux densities [Jy]} & mm & source\\
\cline{2-5}\cline{6-9}\cline{10-13}
$\lambda$ [\micron] & $8.28$ & $12.13$ & $14.65$ & $21.3$ & $12$ & $25$ & $60$ & $100$ & $450$ & $800$ & $1100$ & $1250$ & size [\arcsec] & size [\arcsec] \\
\hline
MM1		& 10.3 & 21.8 & 33.0 & 61.8 & 19.0 & 98.6 & $<$891 & $<$1890 & 49.4 & 7.62 & 1.40 & 1.66 & 21.3 & 18.2\\ 
MM2		& $<$0.05 & $<$0.21 & $<$0.34 & $<$0.73 & $<$0.38 & $<$1.36 & $<$891 & $<$1890 & -- & -- & -- & 0.670 & 18.1 & 14.4\\
MM3		& $<$0.07 & $<$0.26 & $<$0.41 & $<$0.82 & $<$0.61 & $<$2.21 & $<$891 & $<$1890 & -- & -- & -- & 0.261 & 18.0 & 14.2\\ 
\hline
\end{tabular}
\end{center}
\end{table*}
	
In order to derive the spectral energy distribution of the three millimeter peaks in IRAS~18151$-$1208, we retrieve continuum data from the literature. MM1 was imaged for the first time at 450, 800 and 1100~\micron\ by \cite{mccutcheon1995} then at 450 and 850~\micron\ by \cite{williams2004}, and the overall region at 1.2~mm by \citet{beuther2002a}. Uncertainties are homogenized to account for typical calibration errors taken as 20 \% in the millimeter range (1.1 and 1.2~mm) and 30 \% in the submillimeter range (450 and 800~\micron).  In order to compare observations and model, we choose the deconvolved dust emission size $\theta_r$ as a reference for the SED. We deconvolve using $\theta_r = \sqrt{\theta^2_{s}-\theta^2_{mb}}$ where $\theta_s$ is the observed mm-continuum half-peak size derived by \citet{beuther2002a} and $\theta_{mb}$ the HPBW (11\arcsec), using a two-dimensionnal Gaussian fit, assuming that the source is smaller than the beam and has a Gaussian shape. All millimeter ($\lambda > 450$~\micron) peaks $S_\lambda^{peak}$ are then adapted to this reference size using
%

\begin{eqnarray}
S_\lambda = \left(\frac{\theta_{r}}{\theta_{mb}}\right)^{3+p} S_\lambda^{peak}
\label{eqn:fluxadaptation}
\end{eqnarray} 

where $p$ is the density power-law index \citep[derived by][]{beuther2002a} and $S_\lambda$ the adapted flux density used to build the SED. These flux densities are 49.4, 7.62, 1.40 and 1.66~Jy at respectively 450, 850, 1100, 1250~\micron\ for MM1. The process is iterated on MM2 and MM3 and reference sizes with millimeter-continuum peak fluxes are reported in Table~\ref{table2}. Unfortunately the sub-millimeter continuum data for MM2 and MM3 are less abundant than for MM1 as these two sources were not observed by \citet{mccutcheon1995} and \cite{williams2004}.

Concerning mid- and far-infrared flux density measurements, MSX detected MM1 only. Thus we derive a $3\sigma$ noise level inside the beam size around the millimeter-continuum position of MM2 and MM3, taken as an upper limit. We note that the IRAS emission of the region is dominated by MM1 which is confusing upper limits for MM2 and MM3. We therefore improved the IRAS map using the publicly available HiRes maximum correlation method \citep{aumann1990}, showing MM1 detection at 12~\micron\ and 25~\micron. At 60~\micron\ and 100~\micron, a single peak is detected slightly shifted to the west. We conclude that the huge beam width of IRAS sums the three flux densities from the three sources and the environmental filament contribution. Measured flux densities reported in the IRAS catalog at 60~\micron\ and 100~\micron\  are taken as upper limits for the three sources. All results are shown in Table~\ref{table2}.


\section{Modeling the continuum and the molecular emission of MM1 and MM2}
\label{sec:model}

In this section we explain how we derive physical models for MM1 and MM2 and how we use them to investigate the kinematics and the molecular abundances inside the cores. 
Following the method initiated and validated by \citet{vandertak1999,vandertak2000a} we use the dust continuum emission to derive the density and temperature profiles of the cores. 
Dust emission is sensitive to the total column density of dust (optically thin part of the SED in the millimeter wave range), and to the dust temperature and its distribution(mid to far-IR range). 
We try to build the simplest model in order to have the most general representation of the object. Then from the obtained physical model, we can run molecular line modeling to constrain kinematics (line profiles) and chemistry (abundances of observed species). 
Since MM3 is weak or not detected in a number of molecules and does not have a well constrained SED, we do not attempt to model it. We
will restrict ourself to a simple discussion of its evolutionary stage.

\subsection{IRAS~18151$-$1208 MM1}

\subsubsection{Spectral energy distribution}
\label{sssec:sed}

The MM1 core is massive \citep[$M \approx 550$~\msol\ according to][and applying the correction of 1/2 from the associated 2005 erratum]{beuther2002a}. It coincides with the bright IRAS~18151$-$1208 source ($L \approx 2.0 \times 10^4$~\lsol). In first approximation, assuming that the object has a spherical symmetry, we want to derive the best radial distribution of density and temperature that can fit the SED and that takes into account observational constrains.\\

(1) 1D modeling

We use the radiative transfer code MC3D in its 1D version \citep{wolf1999} to calculate the expected dust emission. This code is Monte-Carlo based, and is using the standard MRN \citep{mathis1977} dust grain distribution with the dust opacities from \citet{ossenkopf1994}. The model parameters are the central stellar luminosity $L_*$, its temperature $T_*$, the internal and external radii $r_0$ and $r_{ext}$ of the object, the molecular hydrogen density $n_0$ at $r_0$ and the power-law index $p$ of the density distribution of the form $n(r)=n_0(r/r_0)^p$. 

The bolometric luminosity is derived by integrating the SED, and we obtain $L_{bol}=1.4$~$\times$~10$^4$~\lsol. The temperature of the central star is taken as the corresponding main sequence star, i.e. $T_*=22\,500$~K \citep[type B2V of 10~\msol approximatively, cf.][]{underhill1979}. We adjust $r_{0}$ to fit the dust sublimation radius where $T\simeq 1500$~K. The external radius $r_{ext}$ of the model is derived from the radius of the 1.2~mm continuum emission \citep[here the deconvolved mean of major and minus axe given by][]{beuther2002a} assuming that the source has a Gaussian shape smaller than the beam size. The resulting extension is 18.2\arcsec, hence $r_{ext}=$~27\,300~AU or $\simeq 0.13$~pc at 3~kpc. The power-law index $p$ is set to $-1.2$ according to the continuum profile fit by \citet{beuther2002a}. Finally the only free parameter left is $n_0$ and is derived by fitting optically thin 1.2~mm continuum  emission with a $\chi^2$ calculation. We find $n_0=2.8$~$\times$~10$^{9}$~cm$^{-3}$ at $r_0=21$~AU with $\chi^2=3.94$, hence $\left<n\right>=8.0\times 10^{5}$~cm$^{-3}$ and a total mass of gas equal to 660~\msol. Outer and mean temperatures, formally $T_{ext}$ and $\left<T\right>$, are respectively 25.4~K and 27.3~K. The derived temperature profile can be fitted with a power-law of the form $\log_{10}(T) = \alpha.\log{r}+\beta$ with $\alpha=-0.588$ and $\beta=3.95$. 

\begin{figure}[t!]
   \centering
   \includegraphics[width=\columnwidth]{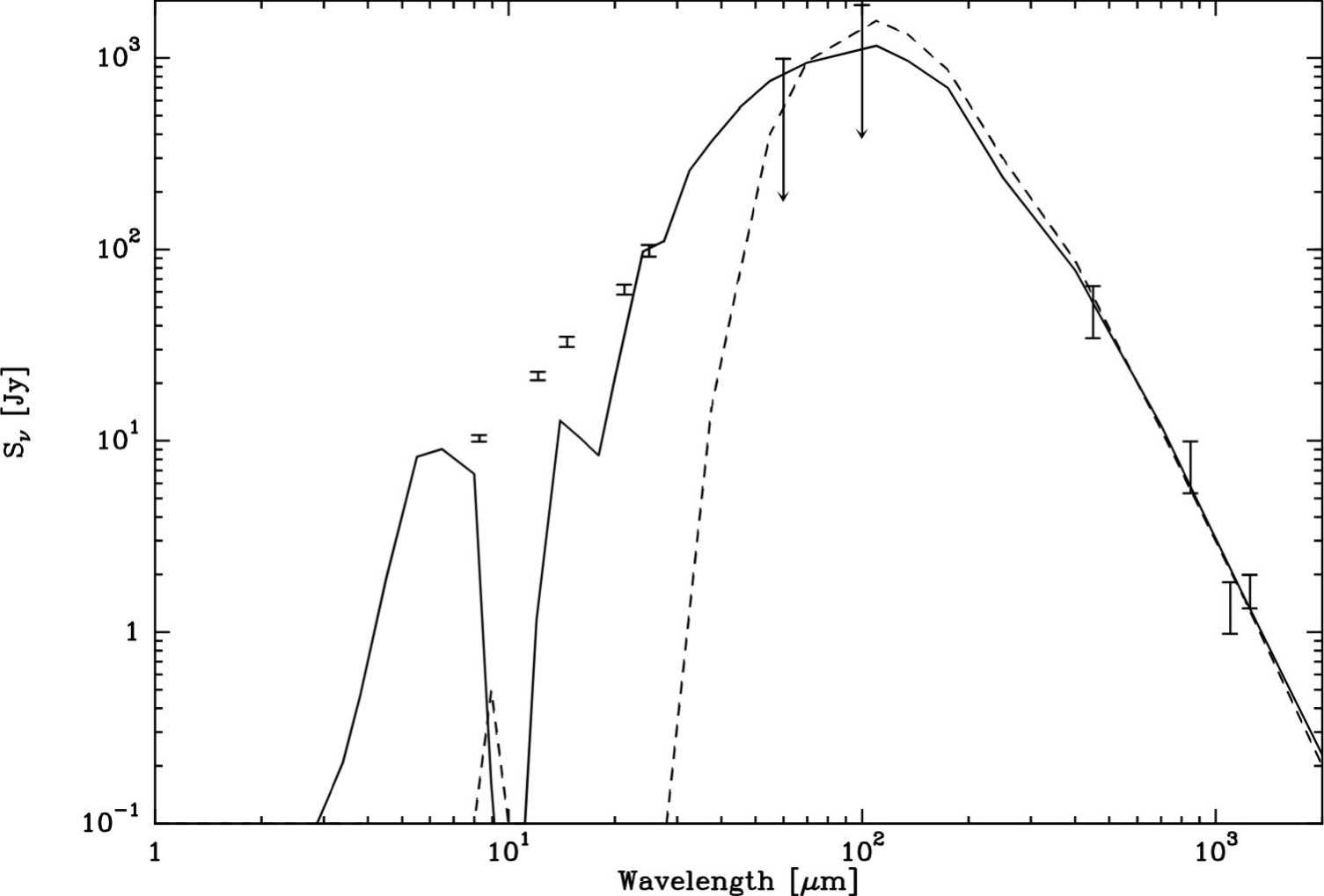}
   \caption{Spectral energy distributions obtained from 1D model (dashed line)
     and 2D model (continuous line) overlaid on observed fluxes of MM1
     source. Dust continuum peak fluxes have been adjusted to fit radial
     extension of the model.  Mid-infrared fluxes are obtained in 2D choosing a
     typical view angle $\theta_{los} \sim 45^{\circ}$ (90$^{\circ}$ = edge-on
     view).}
         \label{seds}
   \end{figure}

The resulting SED (see Fig.~\ref{seds} in dashed line) shows that millimeter and sub-millimeter continuum parts are well modeled. On the
contrary, the mid- and far-infrared emission is not reproduced, due to spherical symmetry that does not permit infrared photons,
coming from inner parts, to leave the source without being reprocessed into photons at millimeter wavelengths. 2D modeling, where
the polar regions contain less matter than in the mid-plane, should therefore reproduce better the observed mid-infrared emission.\\

(2) 2D modeling

We use the 2D version of the MC3D code which was developed for disks. It is
parametrized in radial and altitude directions (r,z) and it assumes the Keplerian disk equilibrium by \cite{shakura1973}:

	\begin{eqnarray}
	n(r,z) = 
	\left\{
	\begin{array}{lr}
	n_0 \left(\frac{r}{r_0}\right)^{p}e^{-\pi c^2 z^2\point \sqrt{2r_{ext}}\point r^{-2.5}} & \mbox{if $r\in[r_0;{r_{ext}}]$}\\
	 & \\
	0 & \mbox{otherwise}
	\end{array}
	\right.
	\label{eqn:keplerian}
	\end{eqnarray}

\begin{figure}[t!]
   \centering
   \includegraphics[width=\columnwidth]{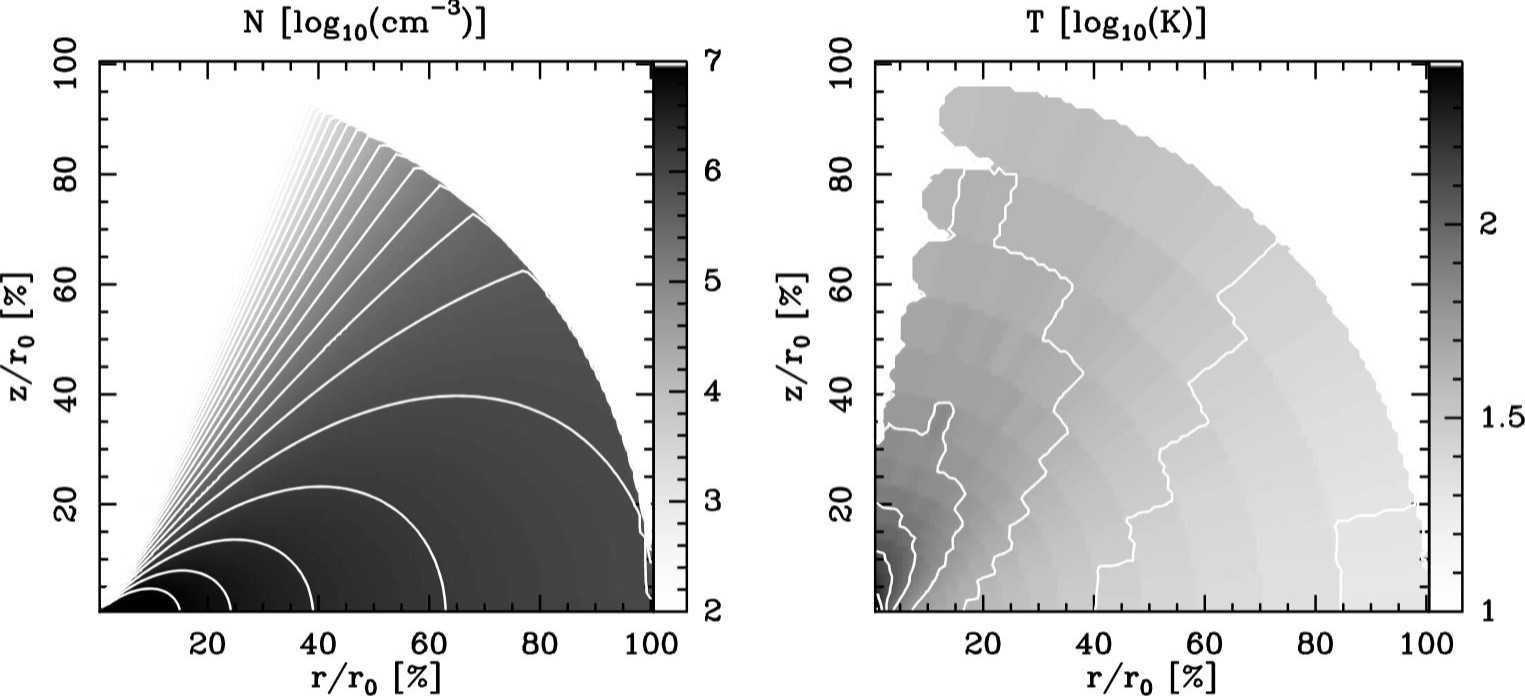}
   \caption{Density and temperature distributions for the 2D model presented in
     a logarithmic grey scale, with 20 logarithmic levels from 10$^2$ to
     10$^7$~cm$^{-3}$ for density and 10 logarithmic levels from 10~K to 263~K.}
         \label{modelmap}
   \end{figure}

        We note that the model has three free parameters: $n_0$ determined as
        for the 1D modeling, the inverse height of scale $c$, and the source
        angle relative to the line of sight $\theta_{los}$. It is mostly
        $\theta_{los}$ which controls the fraction of mid-infrared
        emission. Values close to $45^{\circ}$ can roughly reproduce the SED,
        and we do not further adjust the value to keep the model as general as
        possible. The inverse height of scale $c$ is controlling how thin the
        disk-like structure is. Since MM1 is young and is still mostly a
        spherical core, we try to keep c as small as possible while having
        always a significant mid-infrared emission for typical average lines of
        sight.  Thus $c$ was gradually increased from 0.4 by 0.1 steps until an
        infrared emission greater than 0.1~Jy (limit of detection in the
        observations) could be obtained, finally leading to $c=0.7$. Other
        parameters are kept identical as in 1D model ($T_*$, $L_*$, $r_0$,
        $r_{ext}$ and $p$). Calculations show that the best fit is obtained for
        $n_0=2.7$~$\times$~$10^{9}$~cm$^{-3}$ with $\chi^2=0.732$ (total mass
        $M=830$~\msol). The derived SED is plotted in Fig.~\ref{seds} in
        continuous line.  It illustrates how a 2D modeling can reproduce better
        the observed infrared emission, although the silicate feature is
          not matched. We display the resulting density and temperature
        distributions in Fig.~\ref{modelmap}. One can see that the equatorial
        regions are denser and colder than around the polar axis.  This is why
        mid-IR emission can escape and directly contribute to the SED.


	\subsubsection{CS and C$^{34}$S line emissions}
	
	Since molecular line emission is usually dominated by the mostly
        spherical external layers of the protostellar cores, the 2D description
        might not be required to reproduce molecular lines.
	
	We use the CS and C$^{34}$S line emission to compare the results of
        modeling in 1D and 2D geometries. We directly use outputs of continuum
        models described above, assuming that dust and gas temperatures are
        equal, to which we add parameters for molecular abundances and
        kinematics. The molecular abundances relative to H$_2$, $X$(CS) and
        $X$(C$^{34}$S), mostly determine the line intensities when the opacity
        is low ($\tau\lesssim 1$). Then we use a constant turbulent velocity
        $\varv_T$. It is the main contributor to the line width compared to the
        thermal dispersion. $X$(CS) and $\varv_T$ are therefore the free
        parameters to fit the intensities and widths of the line emission we
        observed. Finally we want to test if an infall motion in our model could
        improve our results, giving at least an upper limit for it. The infall
        velocity is parameterized by the free-fall, power-law distribution
        $\varv_{in}(r) = \varv_{in}\,(r/r_0)^{-0.5}$ \citep{shu1977}. We test
        its effect on modeled line emission by switching between $\varv_{in}=0$
        and $ \varv_{in}=-0.4$~\kms, the lowest value that modifies modeled line
        emission significantly for MM1.
	
	The line profiles are computed with the RATRAN code
        \citep{vandertak2000}. This code allows 1D and 2D modeling using
        respectively multi-shell and grid description of the object. The
        radiative transfer calculation uses the Monte-Carlo method to derive the
        populations of the energy levels and then builds data cubes that are
        convolved with corresponding beam sizes of the telescope to be directly
        compared with the observations.
	
	In the 1D (spherical) geometry, a $\chi^2$ calculation is performed for
        the optically thin C$^{34}$S line ($\tau_{2-1} = 9.0 \times 10^{-2}$ and
        $\tau_{3-2} = 2.2 \times 10^{-1}$) over a grid for
        $f_\mathrm{{C^{34}S}}=X$(C$^{34}$S)$/X$(C$^{34}$S)$_{typ}$ (with
        $X$(C$^{34}$S)$_{typ} = 1.0$~$\times$~$10^{-10}$) and $\varv_T$. The
        best fit is obtained for $(f_\mathrm{{C^{34}S}},\varv_T)=(0.5,1.0)$ with
        $\chi^2=0.956$. Then the grid is refined for $f_\mathrm{{C^{34}S}}$
        only, because model results are less dependent of $\varv_T$ (kept equal
        to $1.0$~\kms). The best result is obtained for $f_\mathrm{{C^{34}S}}
        =0.51$ with $\chi^2=0.952$, leading to the abundance
        $X\mathrm{(C^{34}S)}=5.1~\times~10^{-11}$. Assuming a typical
        [S/$^{34}$S] isotopic ratio of 20 \citep{wilson1994,chin1996,lucas1998}
        the CS emission is modeled with the abundance
        $X\mathrm{(CS)}=1.0~\times~10^{-9}$.
		
\begin{figure*}[t!]
   \centering
   \includegraphics[width=120pt]{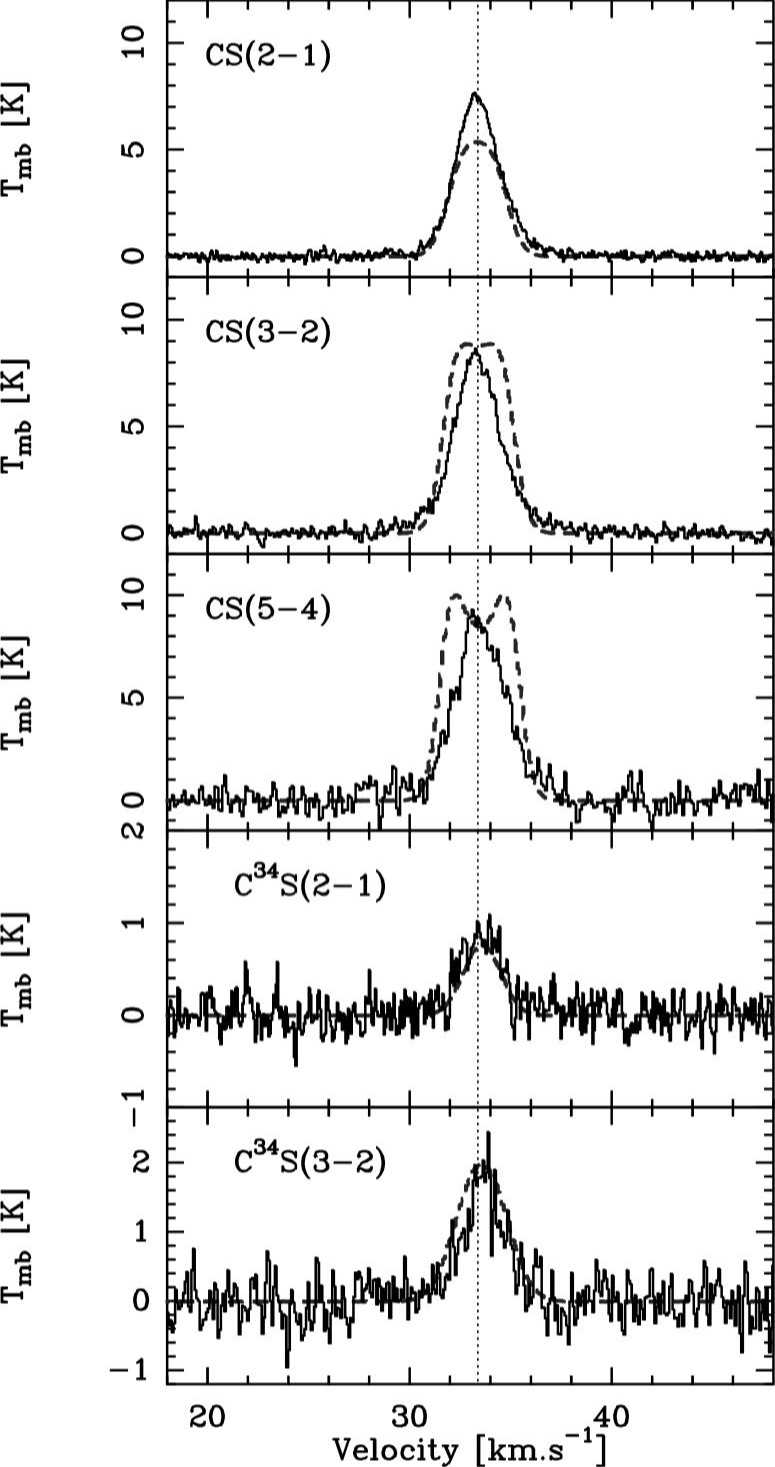}
   \includegraphics[width=120pt]{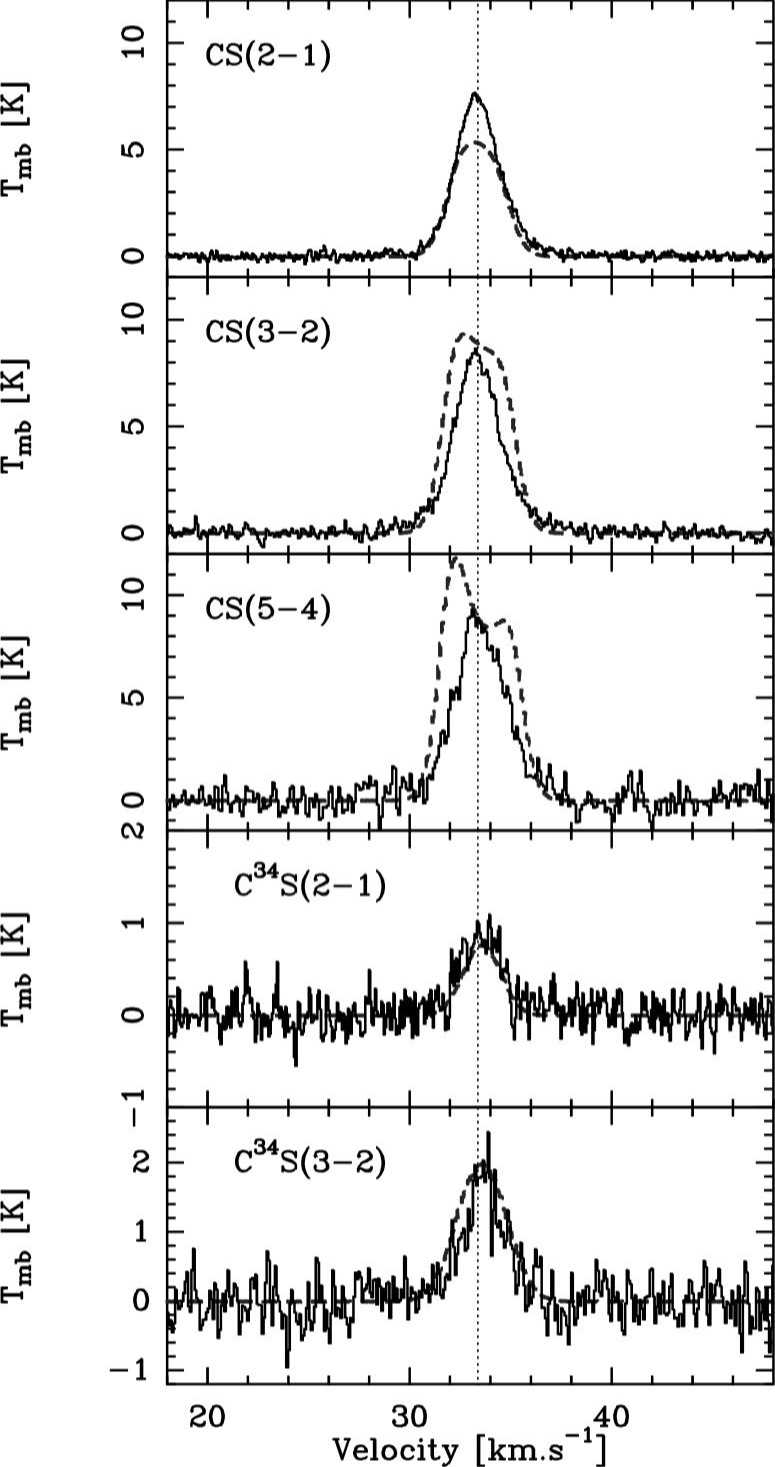} 
   \includegraphics[width=120pt]{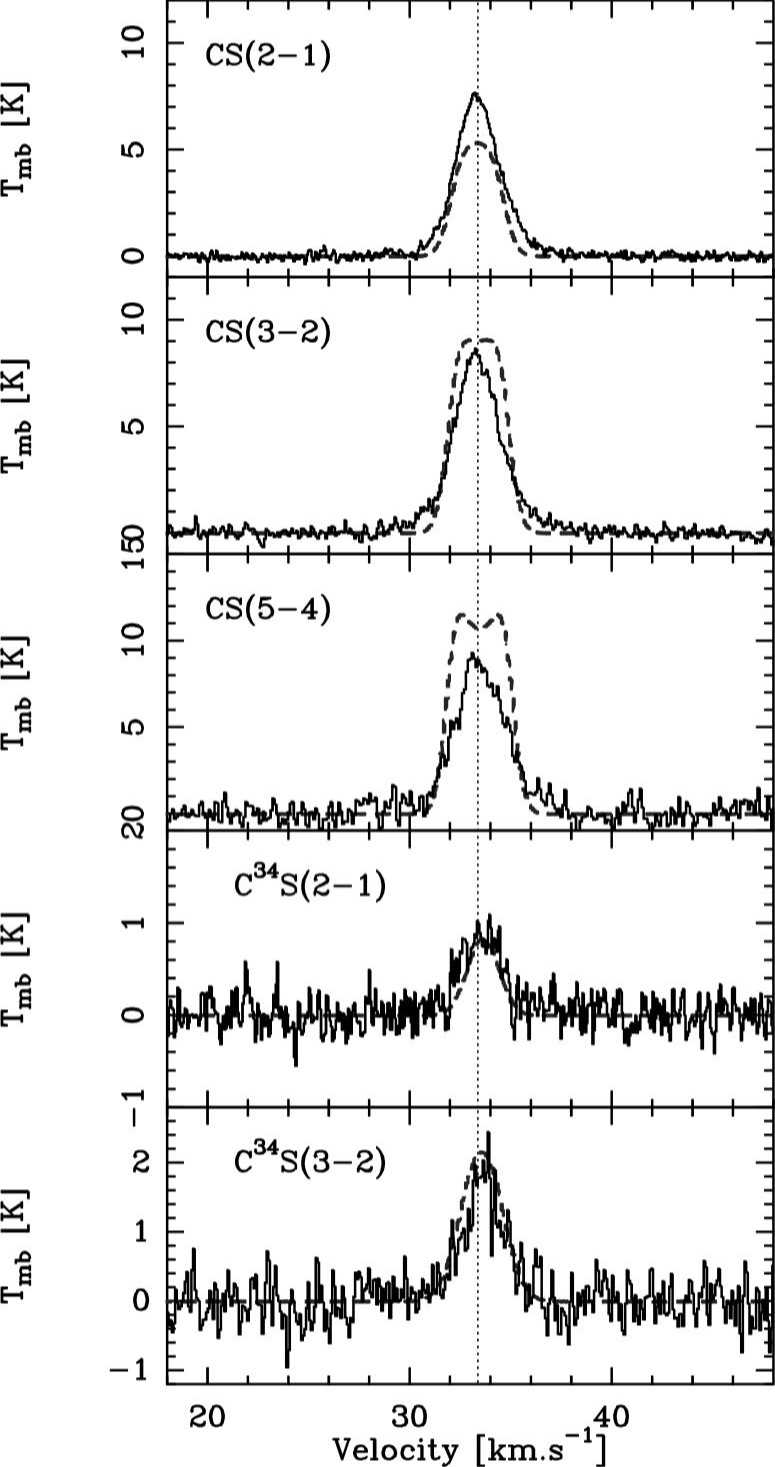}
   \includegraphics[width=120pt]{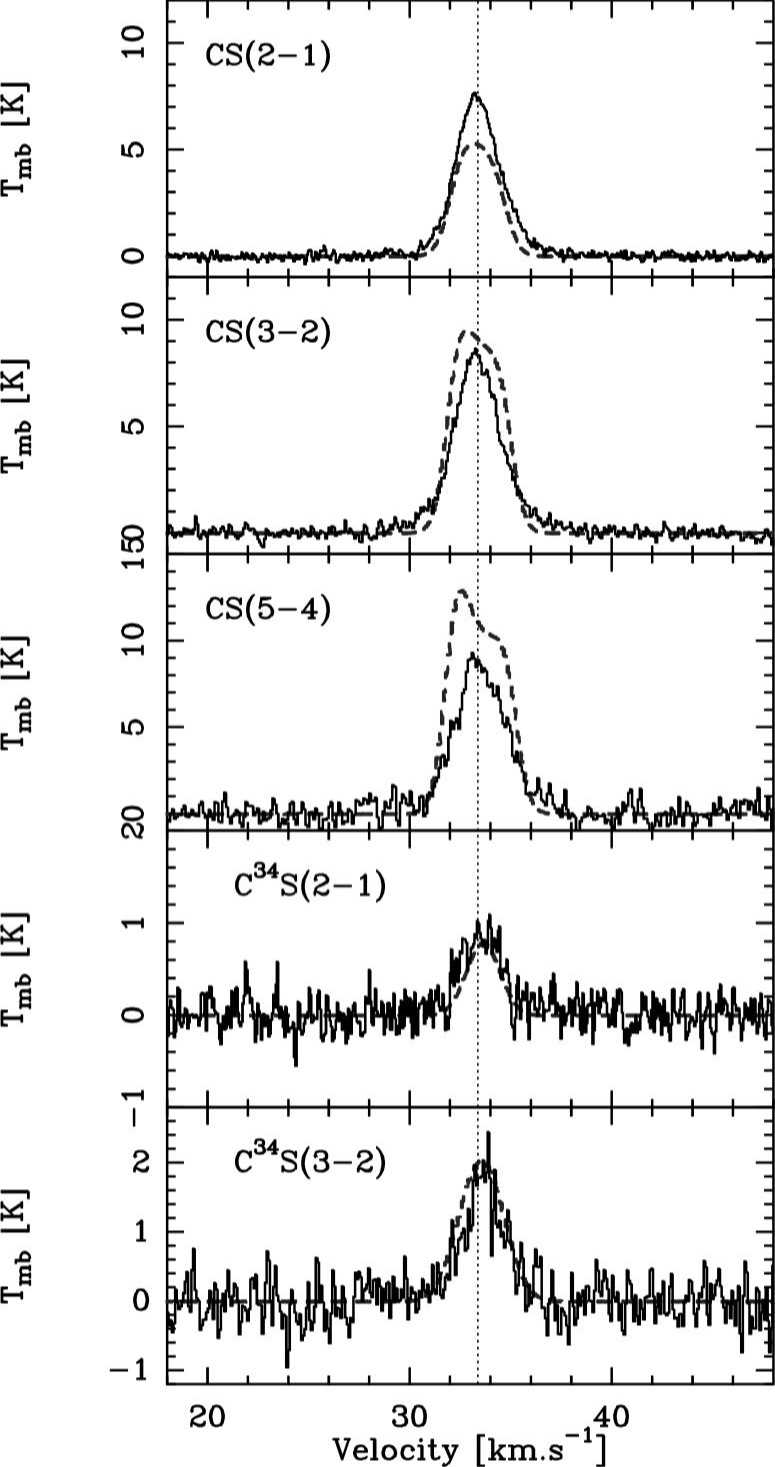}
      \caption{Emission of CS and C$^{34}$S from MM1 object (continuous line) overlaid by --\,from left to right\,-- 1D static, 1D with infall ($\varv_{in}=-0.4$~\kms), 2D static and 2D with infall ($\varv_{in}=-0.4$~\kms) model line emissions (dashed). Spectra velocity resolutions are kept identical to observational resolutions reported in Table~\ref{table4}.}
         \label{fig:csfit-mm1}
   \end{figure*}

	The final results of the line modeling are displayed in Fig.~\ref{fig:csfit-mm1}. Given the very small number (2) of free parameters, the resulting fits are quite
	good. It reproduces well the line intensities. Only the profiles for the higher excitation lines indicate a significant difference between observations and this simple 
	model. There are clear self-absorptions for the \molline{3}{2} and the \molline{5}{4} transitions ($\tau_{5-4} = 3.6$) which indicate an overestimate of large CS column densities
	for the highest excitation regions (central regions). This could well be due to a significant decrease of the CS abundance toward to the center
	of the core (see Sect.~\ref{ssec:disc:cs} for a discussion).
	
	
	We find that a minimum infall velocity of $-0.4$~\kms\ is needed to start to see a clear asymmetric shape of the profile. 
	Introducing infall  improves only slightly the results for the \molline{2}{1} and \molline{3}{2} transitions and makes the fit of the \molline{5}{4} transition worse (cf. Fig.~\ref{fig:csfit-mm1}, second plot from the left). We conclude that an infall velocity in the 1D model does not improve the fits, and we consider as an upper limit the value of $-0.4$~\kms.
	
	For the 2D modeling, the same $\chi^2$ adjustment is performed for $\varv_T$ and $f_\mathrm{{C^{34}S}}$. Interestingly enough, the final best fit is exactly the same than in the 1D geometry with $X\mathrm{(CS)}=1.0~\times~10^{-9}$ and $X\mathrm{(C^{34}S)}=5.1~\times~10^{-11}$. We note that line profiles and relative intensities between the transitions are similar to the 1D case. Only a very slight improvement of the profile of CS\,\molline{5}{4} which is less self-absorbed can be noted. The result is shown on the third plot from the left in Fig~\ref{fig:csfit-mm1}.
	
	Again the effect of an infall velocity on the line profiles is tested. With $\varv_{in}=-0.4$~\kms\ we note that the asymmetric profile of the line emission is less strong than in the 1D case (cf. Fig.~\ref{fig:csfit-mm1}, fourth plot from the left). We however conclude that infall in the 2D case does not improve critically the results.
	
	Comparison between 1D and 2D modeling shows that using a spherical description of the source is sufficient to reproduce CS and C$^{34}$S molecular line
	emission, even if 2D is slightly better with less self-absorption than in the 1D case. In addition we conclude that infall motion is not clearly detected and
	is not necessary to reproduce observations of MM1. We thus decide to use only 1D modeling without infall motion for the other observed lines.
	
	The fact that 2D modeling is not needed for the CS line emission can be explained by the optical depth which is not so high at the corresponding frequencies compared with the mid-IR continuum, or by the asymmetry which is significant only on small scales where the temperature is high (between 300 and 1000~K).
	The observed molecular lines are not tracing these inner regions.
	
	\subsubsection{Other lines: N$_2$H$^+$, HCO$^+$, H$^{13}$CO$^+$ and H$_2$CO}
	\label{sssec:}
	
	We adopt the 1D model derived in the previous section. For each
        molecule, the only free parameter left is therefore its relative
        abundance to H$_2$.\\ 
	
	(1) N$_2$H$^+$ \mollinej{1}{0} modeling

        The N$_2$H$^+$ \mollinej{1}{0} is split by hyperfine structure. 
        Theoretically there are 15 hyperfine components, which blend 
        into seven for sources with low turbulence \citep{caselli95} or into
        three if turbulence is strong, as in our case. In order to correctly fit
        observations from ATNF telescope at Mopra we created the molecular
        datafile for the RATRAN radiative transfer code. Energy levels, Einstein
        coefficients $A_{ul}$ and collisional rates $\gamma_{ul}$ are derived
        from \citet{daniel2004} through the BASECOL database maintained by the
        Observatoire de Paris~\footnote{http://amdpo.obspm.fr/basecol/}.
  
\begin{figure}[t!]
   \centering
   \includegraphics[width=\columnwidth]{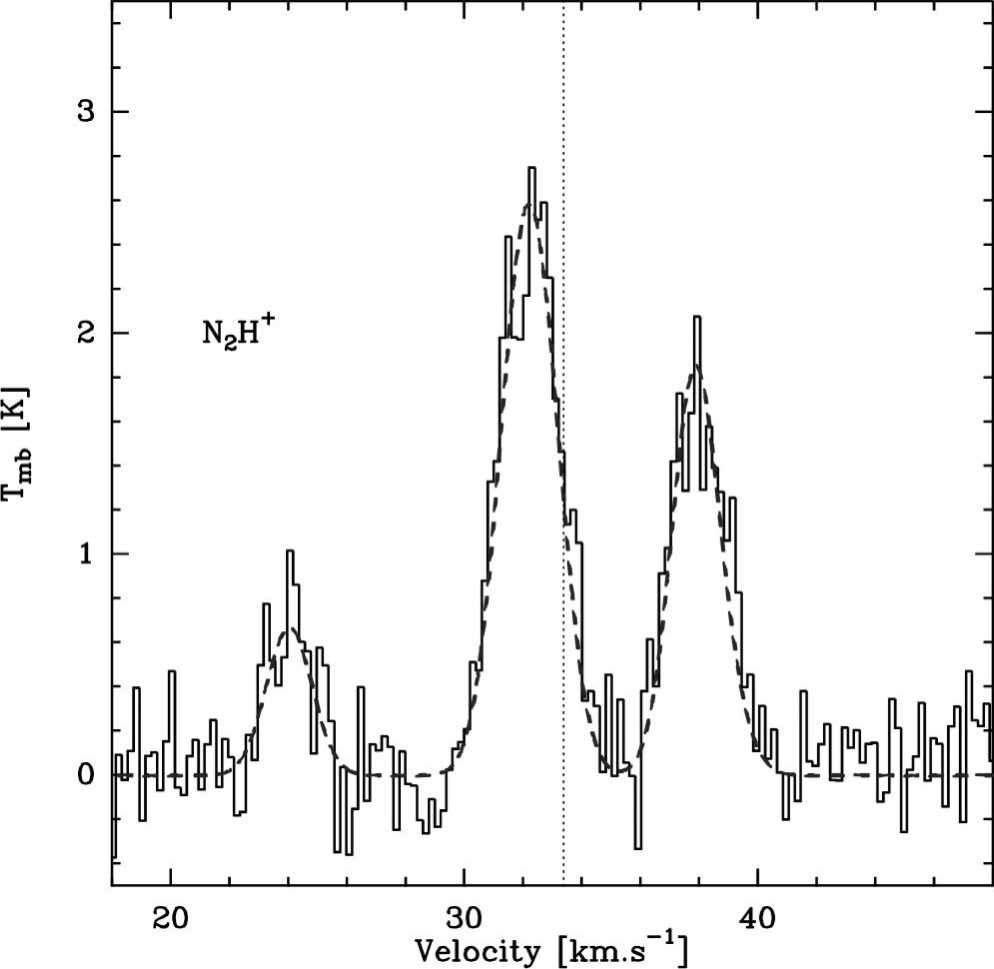}
      \caption{Emission of N$_2$H$^+$ from MM1 object overlaid by 1D model of the source (dashed line). The triplet shape is completely reproduced using our RATRAN molecular datafile. Spectra velocity resolutions are kept identical to observational resolutions reported in Table~\ref{table4}.}
         \label{fig:n2hpp-mm1}
   \end{figure}

        Energy levels included in the molecular datafile vary from $J = 0$ to $J
        = 6$. The hyperfine structure resulting from angular momentum and
        nuclear spin interaction ($F_1$ and $F$ quantum numbers) leads to a
        total of 55 energy levels. The statistical weight of each level is
        determined by \citep[see][]{daniel2005}:
 \begin{eqnarray}
 n_{J,F_1,F} = \frac{(2J+1)(2F+1)}{[J,F_1,F]}
 \end{eqnarray}
 where $[J,F_1,F]$ is the number of magnetic sub-levels for the angular momentum
 $J$. The molecular data file takes into account 119 radiative transitions where
 the first 15 of them fit to the \mollinej{1}{0} transition. The spectra
 obtained are then summed to finally reproduce the whole composite profile of
 the triplet (see Fig.~\ref{fig:n2hpp-mm1}).

 Collision rates $\gamma_{ul}$ reported by \cite{daniel2004} are given for
 helium as collision partner. As described by \cite{schoier2005} we take
 1.4~$\times$~$\gamma_{ul}$ in order to correct for an H$_2$ collision
 partner. Initially collision rates are given for temperatures from 5~K to 50~K
  which is the temperature range of the outer parts of our sources. To be
   able to model the entire sources, we have extrapolated the rates to high
   temperatures using a least-squares method that derives collisional rates from
   a linear fit of the form:
   \begin{eqnarray}
 \log(\gamma_{ul}) = a.log(T) + b
 \end{eqnarray}
 This extrapolation has the advantage of fitting the global trend of the
 collisional rate without any exaggerated values for high temperatures.
  
 Emission of N$_2$H$^+$ is considered as optically thin ($\tau \sim 10^{-2}$)
 due to its typical low abundance. Thus we model it by comparing the velocity
 integrated fluxes of observation and model starting from a typical abundance
 $X\mathrm{(N_2H^+)}=1.0$~$\times$~$10^{-11}$. The ratio of modeled to observed
 fluxes is then used to further correct the abundance while verifying that the
 line profile is always correctly reproduced. Iterating the process gives
 $X\mathrm{(N_2H^+)}=3.5\times 10^{-10}$. An overlay of the model on the
 observation is shown in Fig.~\ref{fig:n2hpp-mm1}.
\\ 
  
\begin{figure}[t!]
   	\centering
   	\includegraphics[width=\columnwidth]{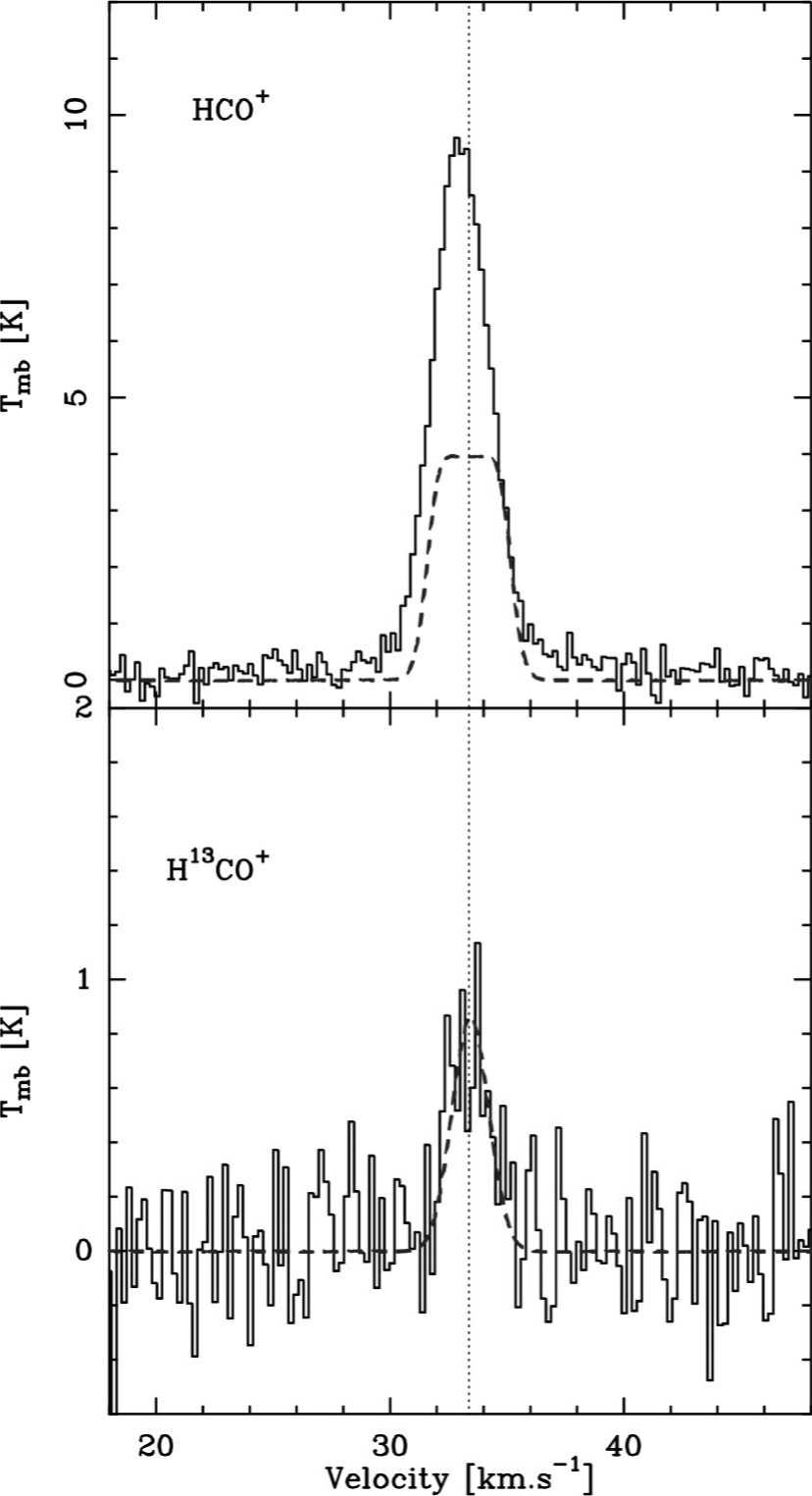}\\
      	\caption{Emission of HCO$^+$ (top) and H$^{13}$CO$^+$ (bottom) \mollinej{1}{0} from MM1 source overlaid by 1D model of the source (dashed line). Resulting abundances are $X\mathrm{(HCO^+)}=2.3$~$\times$~$10^{-9}$ and $X\mathrm{(H^{13}CO^+)}=3.4$~$\times$~$10^{-11}$, according to the typical isotopic ratio $[^{12}$C$]/[^{13}$C$]=67$. Spectra velocity resolutions are kept identical to observational resolutions reported in Table~\ref{table4}.}
         \label{fig:hcop-mm1}
   \end{figure}

   (2) HCO$^+$ and H$^{13}$CO$^+$\,\mollinej{1}{0} modeling\\
   The HCO$^+$ and H$^{13}$CO$^+$ lines are good tracers of high density gas and
   of its kinematics. H$^{13}$CO$^+$\,\mollinej{1}{0} is usually optically thin
   (here we finally get $\tau = 3.4 \times 10^{-2}$) so we adopt the same
   routine to derive molecular abundances as for N$_2$H$^+$. We then get an
   abundance of $X\mathrm{(H^{13}CO^+)}=3.4$~$\times$~$10^{-11}$. Applying a
   typical isotopic ratio $[^{12}$C$]/[^{13}$C$]=67$
   \citep{wilson1994,lucas1998} we model the HCO$^+$ line with
   $X\mathrm{(HCO^+)}=2.3$~$\times$~$10^{-9}$. Results are shown in
   Fig.~\ref{fig:hcop-mm1}.  The resulting modeled HCO$^+$ line is optically
   thick ($\tau = 1.7$) reaching a maximum intensity of $\sim 4$~K. It suggests
   that the region of emission in the model is smaller than the telescope beam
   as expected for a high density tracer. On the other hand the observed
   intensity and profile are very different. It indicates that the bulk of the
   HCO$^+$ rich gas does not follow the general distribution of matter indicated
   by other tracers (dust emission and optically thin lines). The
   HCO$^+$\,\mollinej{1}{0} line seems to have an excess on the blue side which
   could well be associated with the blue-shifted outflow wing which is mostly
   located inside the core in CO (see Fig.~\ref{velocitymaps}).  Generally
   speaking, HCO$^+$ is a good outflow tracer and we speculate that the HCO$^+$
   line is actually dominated by some HCO$^+$ rich gas associated with the
   outflow shocks inside the core ({\it i.e.} at high density as required to
   excite HCO$^+$). This would mean that the derived HCO$^+$ abundance in the
   core envelope
   is only an upper limit and could well be much smaller, except locally in outflow shocks.\\
	
\begin{figure}[t!]
  	\centering
  	\includegraphics[width=\columnwidth]{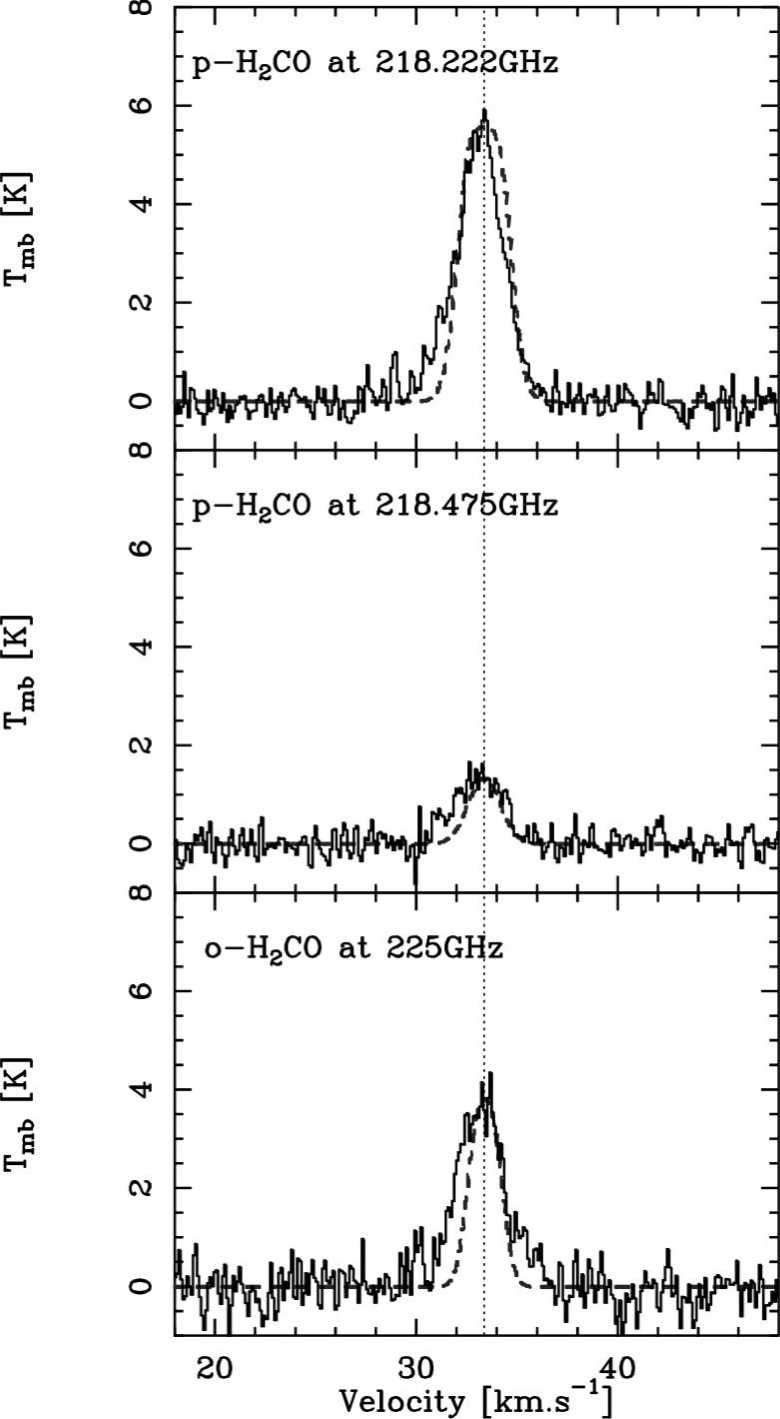}\\
      	\caption{para-H$_2$CO (top and middle) and ortho-H$_2$CO (bottom) emissions from MM1 source overlaid by 1D model of the source (dashed lines). Resulting abundances are $X\mathrm{(para-H_2CO)}=1.5$~$\times$~$10^{-10}$ and $X\mathrm{(ortho-H_2CO)}=1.3$~$\times$~$10^{-10}$, leading to a ratio para/ortho $\simeq 1.2$. Spectra velocity resolutions are kept identical to observational resolutions reported in Table~\ref{table4}.}
         \label{fig:h2co-mm1}
   \end{figure}

   	(3) para- and ortho-H$_2$CO modeling\\
	The ratio between para and ortho populations of H$_2$CO is an interesting parameter in the global context of gas temperature and chemical activity \citep{kahane1984} and can be a useful tool
	to follow chemical evolution of gas inside and between the observed sources. Our modeling indicates that the lines are mostly optically thin ($\tau_{\mathrm{218.2GHz}} = 1.37$,
	$\tau_{\mathrm{218.5GHz}} = 0.09$ and $\tau_{\mathrm{225GHz}} = 0.77$) so we adopt the same routine as described for N$_2$H$^+$. The abundances obtained are as follows
	$X\mathrm{(para-H_2CO)}=1.5$~$\times$~$10^{-10}$ and $X\mathrm{(ortho-H_2CO)}=1.3$~$\times$~$10^{-10}$, leading to a ratio para/ortho $\simeq 1.2$. The agreement between the modeled and
	the observed lines is good. All the lines however tend to show some excess on the blue side, while the ortho line even shows excesses on both sides
	(see Fig.~\ref{fig:h2co-mm1}). In addition we want to emphasize that the two para-H$_2$CO transitions which have two significantly different $E_{\textrm{up}}/k$
	(respectively 21.0~K and 68.1~K) are well reproduced. This is an indication that the excitation temperature of the emitting gas is well reproduced 
	by the physical model.  
	
	\subsection{IRAS~18151$-$1208 MM2}
	

	\subsubsection{Spectral energy distribution}
	\label{sssec:sed-mm2}
	
\begin{figure}[t!]
   \centering
   \includegraphics[width=\columnwidth]{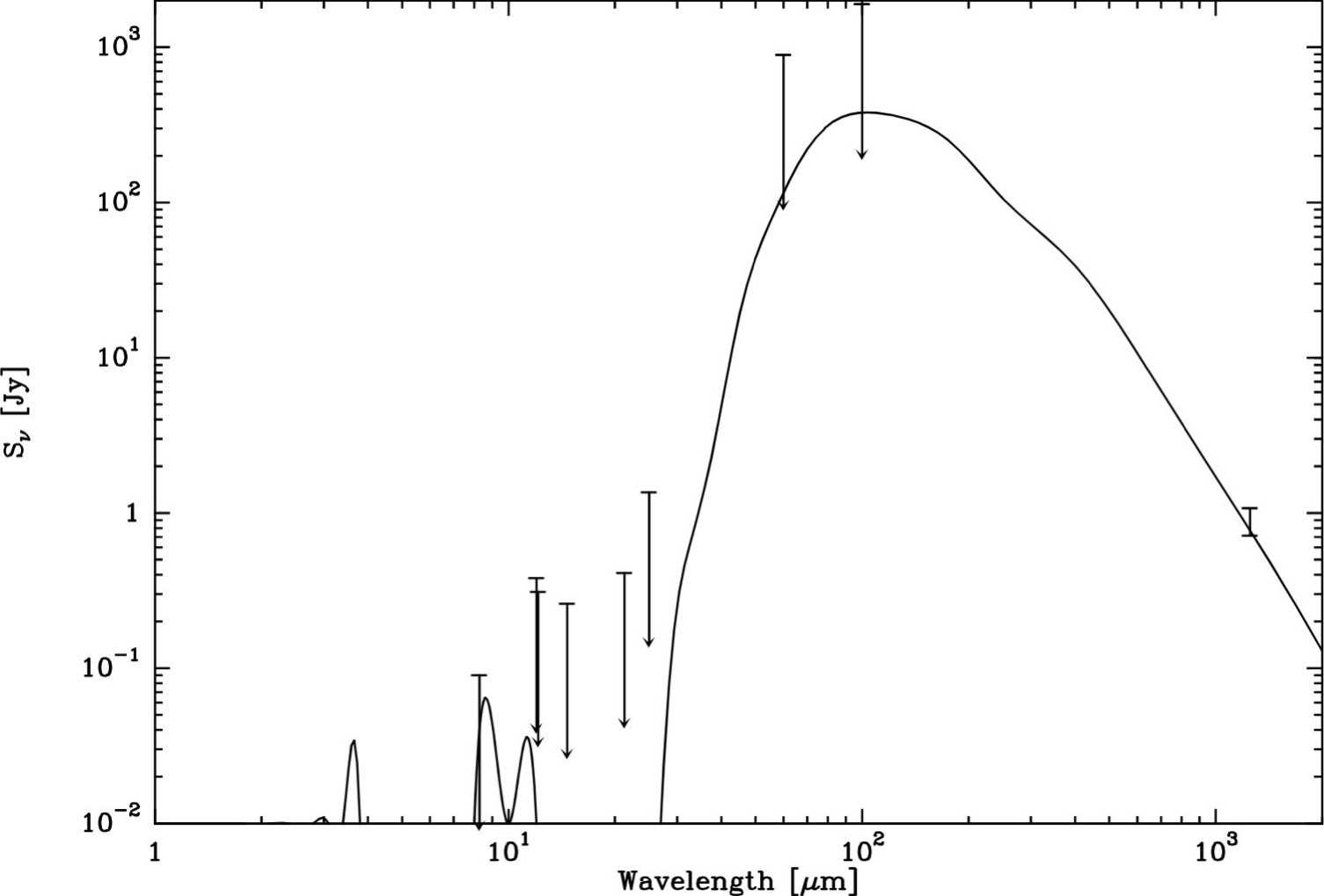}
      \caption{Spectral energy distribution of MM2 overlaid with the 1D model result. Infrared emission has always an upper limit due to the non-detection of the source.}
         \label{fig:sed-mm2}
   \end{figure}

	Since MM2 is not detected in the mid-infrared, and taking into account
        our conclusions from the modeling of MM1, we decide to only model the
        source in a simple 1D, spherical geometry. We cannot derive the
          bolometric luminosity of MM2 by integrating its SED, due to the fact
          that all infrared fluxes are upper limits. Instead, we assume a
          luminosity of 2700~\lsol, which is uncertain by at least a factor of 2.
          This luminosity corresponds to a B4V type star (approximatively 7~\msol) with $T_*
          =16\,600$~K, that we use to describe the heating source of our model.
        The 1.2~mm continuum deconvolved source extension from
        \citet{beuther2002a} is equal to 14.4\arcsec\ thus
        $r_{ext}=$21\,600~A.U. or $\simeq 0.10$~pc at 3~kpc and we assume a
        power-law index $p=-1.3$ from the same paper. Because only one mm-wave
        peak flux density was measured for MM2 (see Table~\ref{table2}), we
        adapt $n_0$ to fit this unique measurement. After two iterations we
        converge to $n_0 = 1.1\times{10^{10}}$~cm$^{-3}$ at $r_0=12.4$~A.U.,
        hence $\left<n\right>=1.1 \times 10^{6}$~cm$^{-3}$ and a total mass gas
        of $M=460$~\msol. We derive $T_{ext}=19.2$~K, $\left<T\right> = 19.4$~K
        and the temperature can be fitted by a power-law of the form
        $\log_{10}(T) = \alpha.\log{r}+\beta$ with $\alpha=-0.614$ and
        $\beta=3.85$.  The resulting SED does not show any emission in the mid-
        and far-infrared, in agreement with the observations (see
        Fig.~\ref{fig:sed-mm2}) and as expected for a simple 1D description.
	
	\subsubsection{Molecular line emission modeling}
	\label{sssec:mm2-model}
	
	We use the same procedure to model the molecular lines toward MM2 as for MM1. We therefore have derived the molecular abundance relative to H$_2$, and the turbulent 
	velocity $\varv_T$. We have also tested a possible infall velocity field of the form $\varv_{in}(r) = \varv_{in}\,(r/r_0)^{-0.5}$.
	
	Emission of optically thin C$^{34}$S lines ($\tau_{2-1}= 4.4\times 10^{-2}$ and $\tau_{3-2}= 4.4\times 10^{-2}$) in MM2 object are treated by taking two different grids for the
	$\chi^2$ calculation over its intensity and area. The first grid is $(f_\mathrm{{C^{34}S}},\varv_T)=([0.1,0.2 ... 0.5],[1.5,1.6 ... 2.0])$ and shows a best fit for
	$(f,\varv_T)=(0.3,1.7)$, the second one is finer with $f_\mathrm{{C^{34}S}}=[0.25,0.26 ... 0.35]$ and a fixed $\varv_T=1.7$~\kms\ because of the low dependence of the modeled line
	on this parameter, compared to $f_\mathrm{{C^{34}S}}$. It gives the best fit for $f_\mathrm{{C^{34}S}}=0.27$, hence an abundance $X\mathrm{(C^{34}S)}=2.7$~$\times$~$10^{-11}$.
	As for MM1, the resulting modeled lines reproduce well all transitions except the CS\,\molline{5}{4} line which is heavily overestimated in intensity. Again it points to
	a lower abundance of CS in the inner regions compared to the outside. 

\begin{figure}[t!]
   \centering
   \includegraphics[width=120pt]{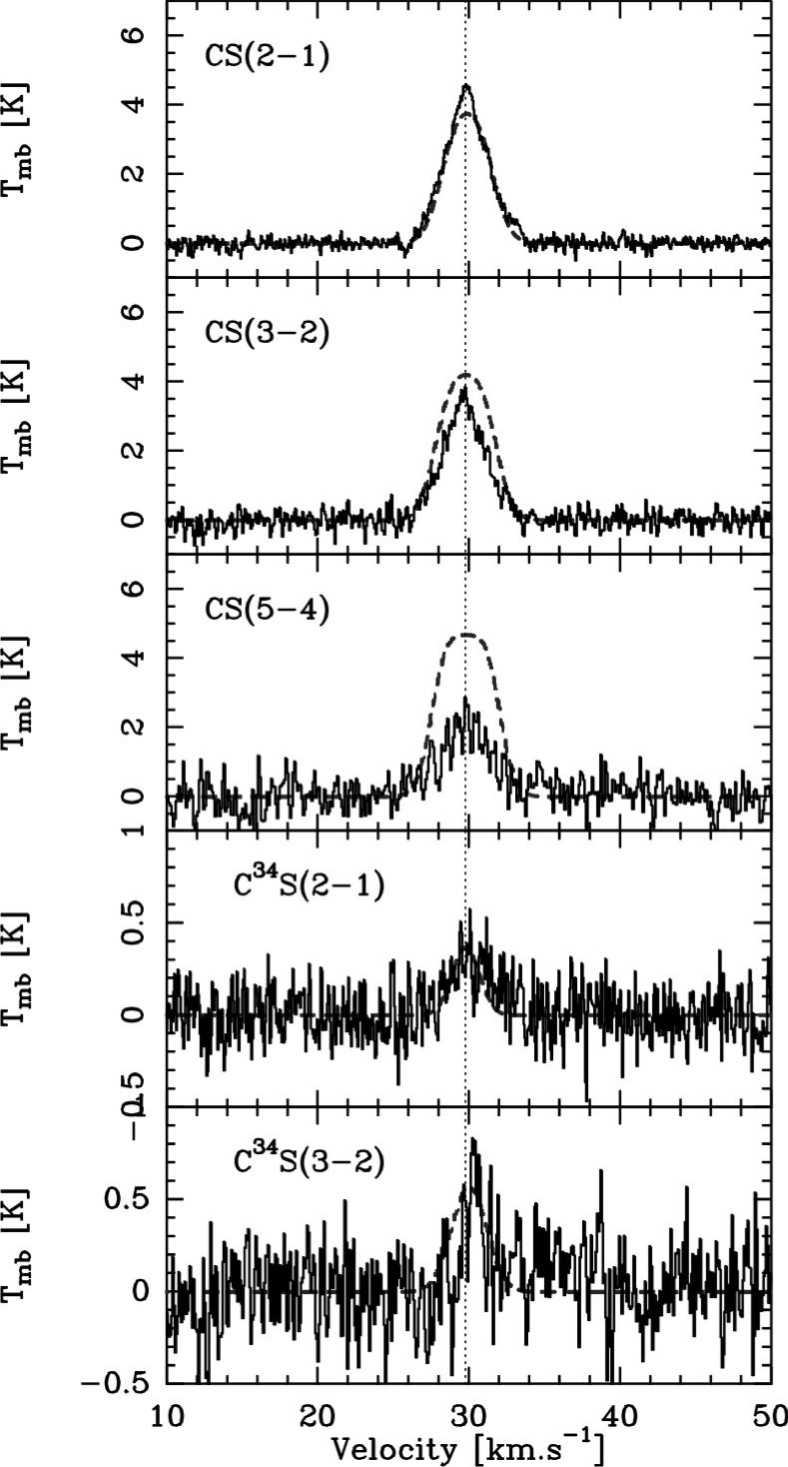}
   \includegraphics[width=120pt]{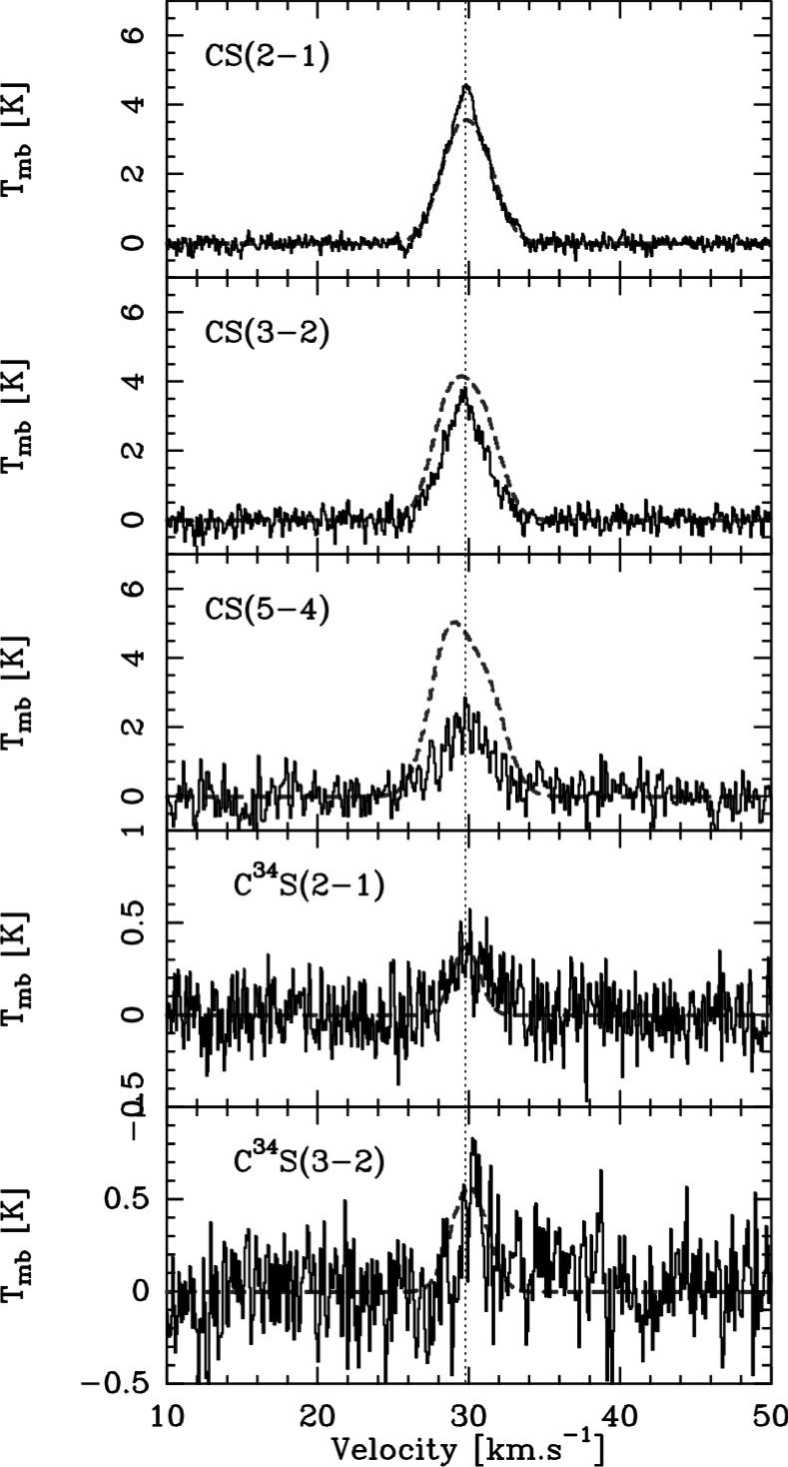}
      \caption{Emission of CS from MM2 object overlaid by 1D static (left) and 1D with infall ($\varv_{in} = -1.0$~\kms) model of the source (dashed lines). Spectra velocity resolutions are kept identical to observational resolutions reported in Table~\ref{table4}.}
         \label{fig:csfit-mm2}
   \end{figure}

	 Then, as above for MM1, we add an infall velocity profile following a typical $r^{-0.5}$ distribution. The asymmetric shape is obtained for a minimum infall velocity
	 $\varv_{in} =-1.0$~\kms. The resulting profile is unchanged for \molline{2}{1} transition, is slightly improved for the \molline{3}{2} but shows an excess of blue emission which is
	 not observed for CS\,\molline{5}{4} (see Fig.~\ref{fig:csfit-mm2}, right). Therefore including infall velocity distribution does not
	 improve our fit and hence does not seem to be necessary. It gives an upper limit for the infall velocity of $\varv_{in} =-1.0$~\kms.

\begin{figure}[t!]
   \centering
   \includegraphics[width=\columnwidth]{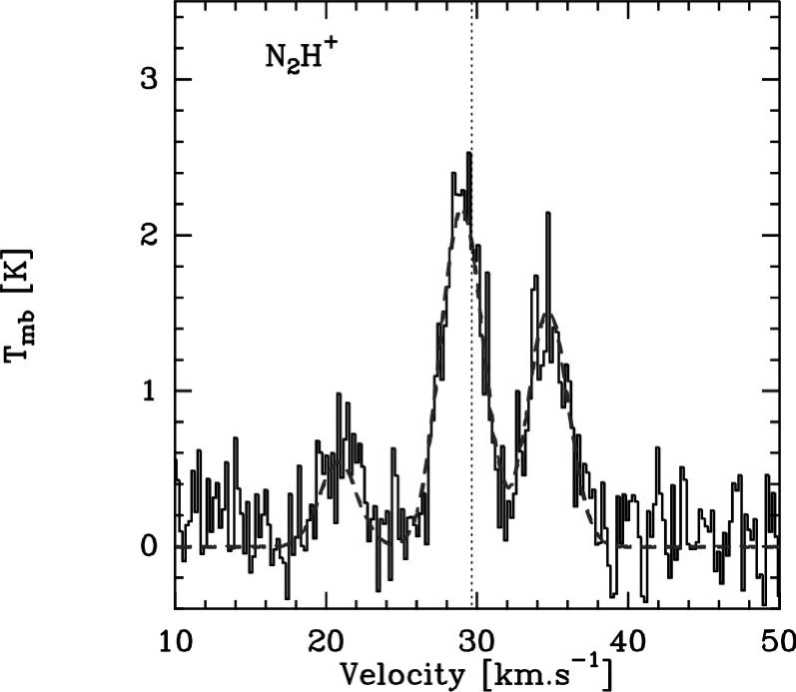}
      \caption{Line emission of N$_2$H$^+$ from MM2 object overlaid by the 1D model of the source (dashed line). Spectra velocity resolutions are kept identical to the observational resolutions reported in Table~\ref{table4}.}
         \label{fig:n2hpp-mm2}
   \end{figure}

	The optically thin emission of N$_2$H$^+$ ($\tau \sim 10^{-2}$) in MM2 is treated by comparing the observed and modeled velocity integrated area and using it to adapt an initial
	typical abundance ($1.0\times 10^{-10}$). The resulting fit for N$_2$H$^+$ is shown in Fig.~\ref{fig:n2hpp-mm2} and corresponds to $X\mathrm{(N_2H^+)}=6.3$~$\times$~$10^{-10}$.
        As for MM1, the quality of the fit is high.

\begin{figure}[t!]
   	\centering
   	\includegraphics[width=\columnwidth]{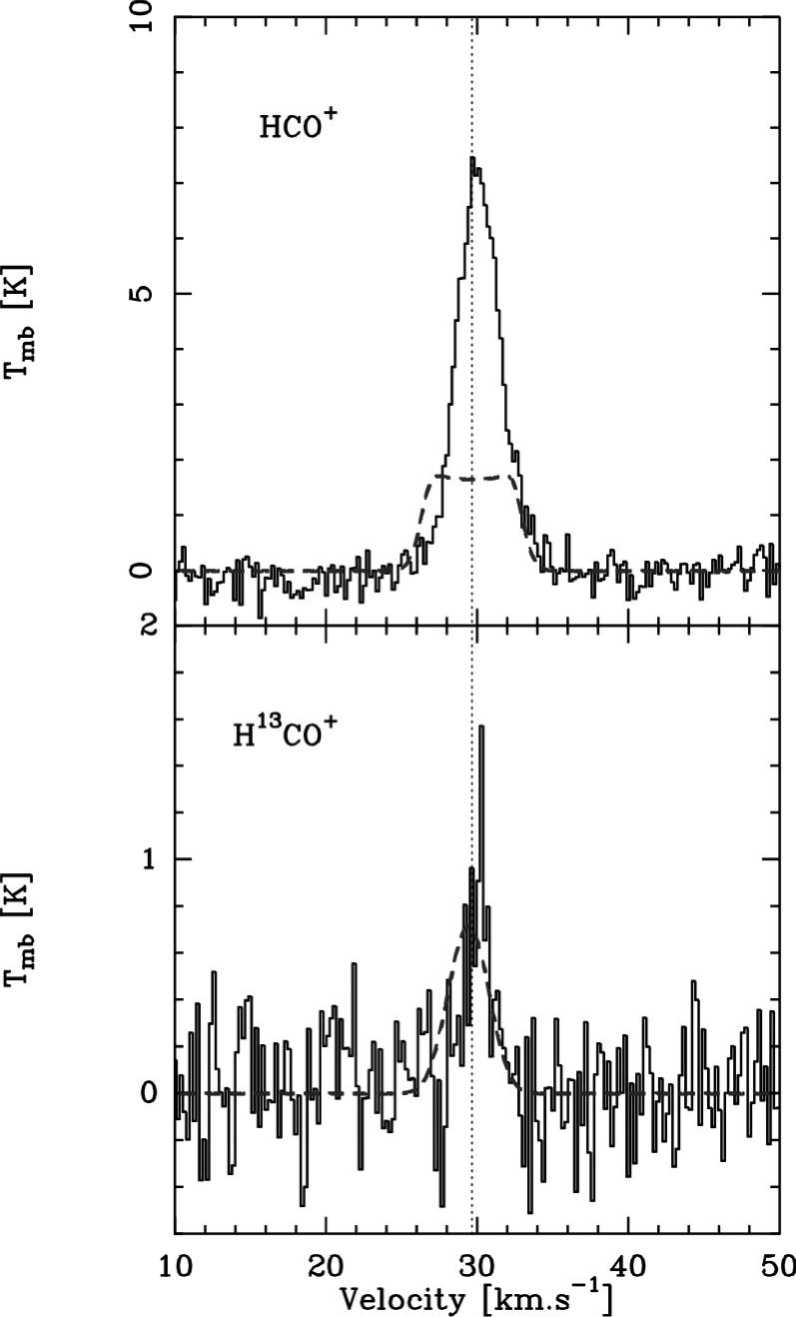}\\
      	\caption{Emissions of HCO$^+$ (top) and H$^{13}$CO$^+$ (bottom) ($J$=1--0) from MM2 source overlaid by 1D model of the source (dashed line). Spectra velocity resolutions are kept identical to observational resolutions reported in Table~\ref{table4}.}
         \label{fig:hcop-mm2}
   \end{figure}

	For HCO$^+$ and H$^{13}$CO$^+$ we obtain the following abundances: $X\mathrm{(HCO^+)}=5.1\times 10^{-9}$ and $X\mathrm{(H^{13}CO^+)}=7.6\times 10^{-11}$. 
	The best fit result is shown in Fig.~\ref{fig:hcop-mm2}. As for MM1, the modeled HCO$^+$ line emission does not reproduce well the observations. 
	The HCO$^+$ column density derived from the H$^{13}$CO$^+$ line implies an heavily saturated ($\tau = 19$) HCO$^+$ line which is not observed. Obviously in contrast
	to all the other molecules, the distribution of HCO$^+$ does not follow the simple spherical geometry of the model.
	
\begin{figure}[t!]
  	\centering
  	\includegraphics[width=\columnwidth]{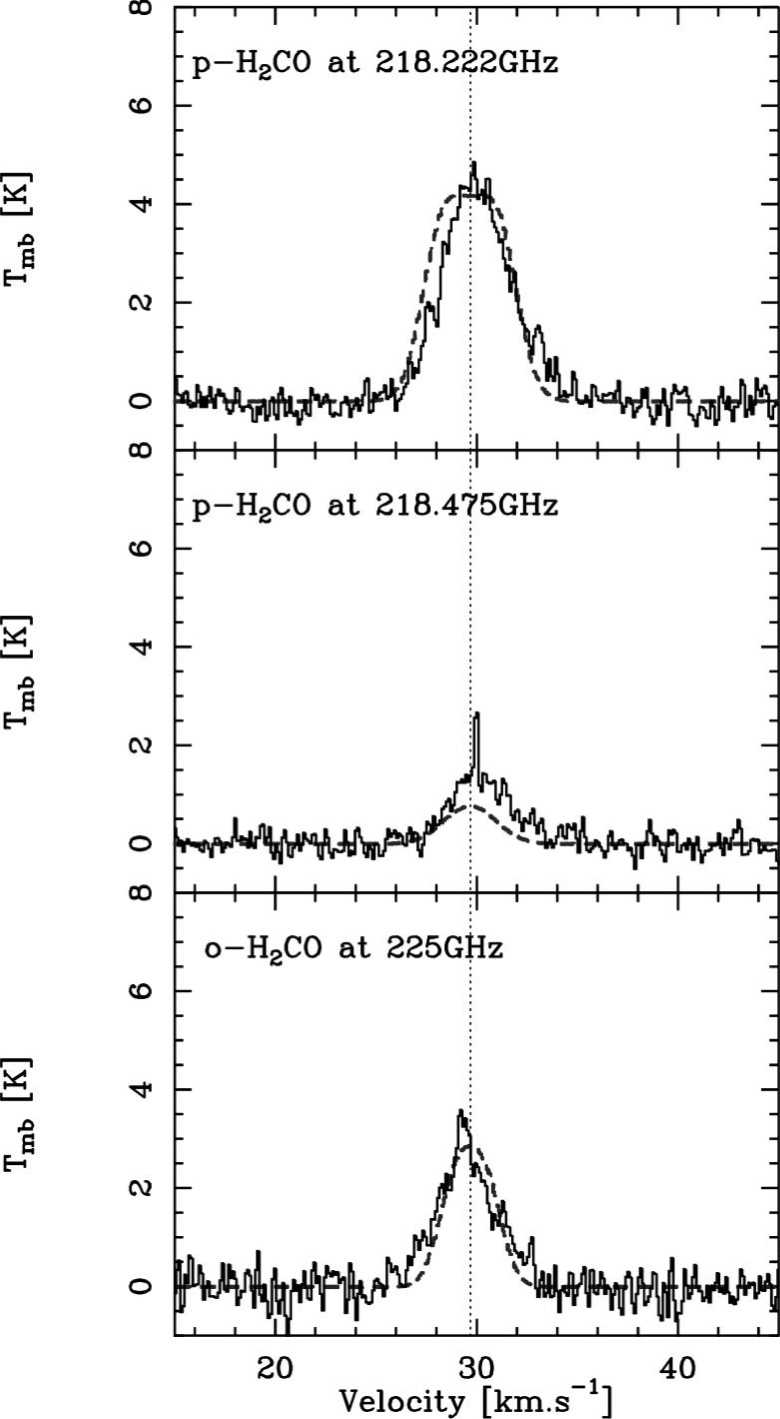}\\
      	\caption{para-H$_2$CO (top and center) and ortho-H$_2$CO (bottom) emissions from MM2 source overlaid by 1D model of the source (dashed lines). Spectra velocity resolutions are kept identical to observational resolutions reported in Table~\ref{table4}.}
         \label{fig:h2co-mm2}
   \end{figure}

	We then derive the best abundances for the para and ortho-H$_2$CO to reproduce the three observed transitions (see Fig.~\ref{fig:h2co-mm2}).
        The abundance obtained for the ortho transition at 225~GHz is $X\mathrm{(ortho-H_2CO)}=2.0$~$\times$~$10^{-10}$. For the two para transitions, no unique
	abundance can be derived. For the best fit of the 218.222~GHz line emission, the abundance obtained is $X\mathrm{(para-H_2CO)}=2.6$~$\times$~$10^{-10}$ but the 
	218.475~GHz line is then too weak by a factor of almost 2; see Fig.~\ref{fig:h2co-mm2}. At the other extreme,
	if the 218.475~GHz line is fitted, the abundance obtained is $X\mathrm{(para-H_2CO)}=4.8\times 10^{-10}$. In contrast to MM1 for which
	a correct ratio of the two para lines was obtained, the modeled ratio (218.222 over 218.475) for MM2 appears to be too large. 
	Since the weaker 218.475~GHz line emission has a higher upper energy level 
	($E_{\textrm{up}}/k=68.1$~K) it suggests that the emitting gas is actually warmer in MM2 than what is represented by the simple 1D model.
	The resulting para to ortho ratio is therefore in the range $\simeq 1.3$-- 2.4.
	
\section{Discussion}

\begin{table}
\begin{center}
	\begin{tabular}{l|cc}
	~ 					& MM1 							& MM2 \\
	\hline
	\hline
	$L$					& 14\,000~\lsol						& 2\,700~\lsol \\
	$T_*$					& 22\,500~K						& 16\,600~K \\
	$M_{gas}$				& $660\pm 130$~\msol						& $460 \pm 100$~\msol \\
	$r_{ext}$				& 27\,300~A.U.						& 21\,600~A.U. \\
	$n_0$				& $2.8\,(0.6)$~$\times$~10$^{9}$~cm$^{-3}$		& $1.1\,(0.2)$~$\times$~10$^{10}$~cm$^{-3}$ \\
	$\left<n\right>$			& $8.0\,(2.0)$~$\times$~10$^{5}$~cm$^{-3}$		& $1.1\,(0.2)$~$\times$~10$^{6}$~cm$^{-3}$ \\
	$p$						& -1.2~$^a$					& -1.3~$^a$ \\
	$T_{ext}$				& $25.4\,(0.5)$~K			& 19.2\,(0.4)~K \\
	$\left<T\right>$			& $27.3\,(0.2)$~K			& 19.4\,(0.2)~K \\
	$\alpha$					& $-0.588\,(0.011)$			& -0.614\,(0.012) \\
	$\beta$					& $3.95\,(0.03)$				& $3.85\,(0.04)$ \\
	$r_{in}$					& 21\,(2)~A.U.				& 12\,(1)~A.U. \\
	\hline
	\hline
	$\varv_T$			& $1.0\,(0.1)$~\kms					& $1.7\,(0.1)$~\kms \\
	$X$(CS)				& $1.0\,(0.3)$~$\times$~10$^{-9}$	& $5.4\,(1.5)$~$\times$~10$^{-10}$ \\
	$X$(C$^{34}$S)		& $5.1\,(1.4)$~$\times$~10$^{-11}$	& $2.7\,(0.8)$~$\times$~10$^{-11}$\\
	$X$(N$_2$H$^+$)		& 3.5\,(0.8)~$\times$~10$^{-10}$		& 6.3\,(1.4)~$\times$~10$^{-10}$\\
	$X$(HCO$^+$)			& 2.3\,(0.7)~$\times$~10$^{-9}$		& 5.1\,(1.6)~$\times$~10$^{-9}$\\
	$X$(H$^{13}$CO$^+$) 	& 3.4\,(1.1)~$\times$~10$^{-11}$		& 7.6\,(2.4)~$\times$~10$^{-11}$\\
	$X$(p-H$_2$CO)	& 1.5\,(0.4)~$\times$~10$^{-10}$		& 2.6\,(0.7)--4.8(1.3)~$\times$~10$^{-10}$\\
	$X$(o-H$_2$CO)	& 1.3\,(0.4)~$\times$~10$^{-10}$		& 2.0\,(0.6)~$\times$~10$^{-10}$\\
	$$[para/ortho]  		& $\simeq$ 1.2						& $\simeq$ 1.3--2.4 \\
	\hline
	\multicolumn{3}{c}{} \\
	\end{tabular}\\
	$^a$ Values from continuum map analysis by \citet{beuther2002a}.
\end{center}
\caption{Summary of results from dust continuum emission modeling (above mid-line) and from molecular line emission modeling (below mid-line).
The relative uncertainties on the derived parameters are given inside the parentheses and are at the 3$\,\sigma$ level. They correspond
to the rms noise propagated through the modeling process. The additional
absolute uncertainty on masses and abundances, mostly due to the uncertain emissivity of dust, is given in Sect.~\ref{ssec:disc:dens-prof}.}
\label{tab:summary}
\end{table}

\subsection{MM2: a new massive protostar driving a powerful outflow.}

MM2 is a millimeter source without infrared counterpart (MSX or IRAS). It is
even seen in absorption over the local background in the MSX 8\micron\ image
(see Fig.~\ref{fig:msx-map}). A water maser has been detected toward the core by
\citet{beuther2002c}.  With the addition of our newly discovered CO outflow
driven by MM2, it all clearly points to a protostellar nature of MM2.  Inside
the core size of 0.22~pc (14.4\arcsec, see Table~\ref{ssec:sed}) a total mass of
460~\msol is obtained with MC3D (dust emissivity for the MRN grain distribution;
see Table~\ref{tab:summary}).  MM2 is therefore a newly discovered 
massive protostellar core which is not detected in the mid- or far-IR: an
IR-quiet massive protostellar object.  A more complete comparison with the
detailed analysis obtained for low-mass protostars and in Cygnus X by
\citet{motte2007} is given in the following section.
	
\subsection{Evolutionary stages of MM1, MM2 and MM3.}

In contrast to low-mass star formation \citep[\textit{e.g.}][]{andre2000}, the
evolutionary sequence for high-mass stars is not well constrained and
understood. The general lack of spatial resolution leads to discussions of
evolutionary sequences mostly applied to massive clumps (such as HMPOs, typical
size of 0.5$\,$pc) or to high-mass dense cores (typical size of 0.1$\,$pc) which
cannot usually be directly compared to protostars which have physical sizes more
of the order of $0.01-0.05\,$pc \citep[\textit{e.g.}][]{motte1998, motte2001}.
A recent attempt to clarify these different scales and associated evolutionary
sequences is compiled in \citet{motte2007}; see their Table~4.  From the
complete survey of the relatively nearby Cygnus X complex, a total of 40
high-mass dense cores were recognized with an average size of 0.13~pc, similar
to sizes of nearby, low-mass dense cores but 20 times more massive and 5 times
denser. These high-density cores were all found to be already protostellar in
nature and were proposed to well represent the earliest phases of high-mass star
formation. Among the 40 cores, 15 were already associated with an UC-H{\sc{ii}}
region and bright in the infrared, 8 were only bright in the infrared, and 17
were IR-quiet protostellar cores. This sequence is proposed to be mostly an
evolutionary sequence, the IR-quiet cores being the precursors of the IR-bright
cores, and then of UC-H{\sc{ii}} regions. We will discuss the cores in
IRAS~18151$-$1208 in the framework of these recent Cygnus X results.
	
First of all no bright radio source is detected in the whole IRAS~18151$-$1208
clump. All the massive protostellar objects detected in the clump are
necessarily objects in an evolutionary stage prior to the formation of an
UC-H{\sc{ii}} region.  In Fig.~\ref{fig:msx-map}, the distribution of
mid-infrared emission from MSX is displayed.  It is clear that the main bright
infrared source is associated with MM1 while the MM2 area is devoid of
mid-infrared emission. The core is actually even seen in absorption over the
local background. MM3 is more complicated since a moderately bright MSX source
is situated close to the center of the core.
	
MM1 has a mass of $660$~\msol in a radius of 27\,300~A.U. (core size of 0.27~pc)
obtained from the fit of the SED with MC3D (Table~\ref{tab:summary}). It has a
bolometric luminosity higher than 10$^4$~\lsol, and is bright in the mid-IR
(S$_{20\,\upmu\rm{m}} = 61$~Jy). It drives a powerful bipolar outflow and
therefore hosts at least one mid-IR bright massive protostar. Using the same
dust properties (millimeter dust emissivity and average temperature) as in
\citet{motte2007}, and inside a 0.13~pc size, the MM1 core would have a mass as
high as 74~\msol, and would therefore be among the most massive cores of Cygnus
X.  With an equivalent S$_{20\,\upmu\rm{m}} = 190$~Jy at 1.7~kpc (distance of
Cygnus X), it would be the brightest core which is not associated with an
UC-H{\sc{ii}} region \citep[see Fig.~7 in ][]{motte2007}.
It is however not as bright in the mid-IR as the well-known AFGL~2591 source
described by \citet{vandertak1999}.  In the \citet{motte2007} classification,
MM1 is therefore a high-luminosity IR (or IR bright) protostar.
	
Except its non detection in the mid- and far-IR (S$_{20\,\upmu\rm{m}} \lesssim
2$~Jy), the properties of MM2 ressemble those of MM1.  It has a mass of
$460$~\msol in a radius of 21\,600~A.U. (core size of 0.22~pc) obtained with
MC3D (Table~\ref{tab:summary}), a still high bolometric luminosity of
2700~\lsol, and it drives a powerful bipolar outflow. Using the same dust
properties as in \citet{motte2007}, and inside a 0.13~pc, the MM2 core would
have a mass of 45~\msol, and would therefore be among the 40 high-density cores
of Cygnus X. In the \citet{motte2007} classification, MM2 is a mid-IR-quiet
massive protostar.
	
Observations of MM3 does not permit to conclude definitively on its
nature. Although it is massive with a flatter density profile (about 200~\msol
and $p=0.8$, \cite[][]{beuther2002a} and no outflow detected, any mid-IR
emission is confused due to the nearing confusing source detected by MSX and the
presence of a bright IR filament in the background.  At best one can only
propose that MM3 is a probable prestellar core.
	
\begin{figure}[t!]
   \centering
   \includegraphics[width=\columnwidth]{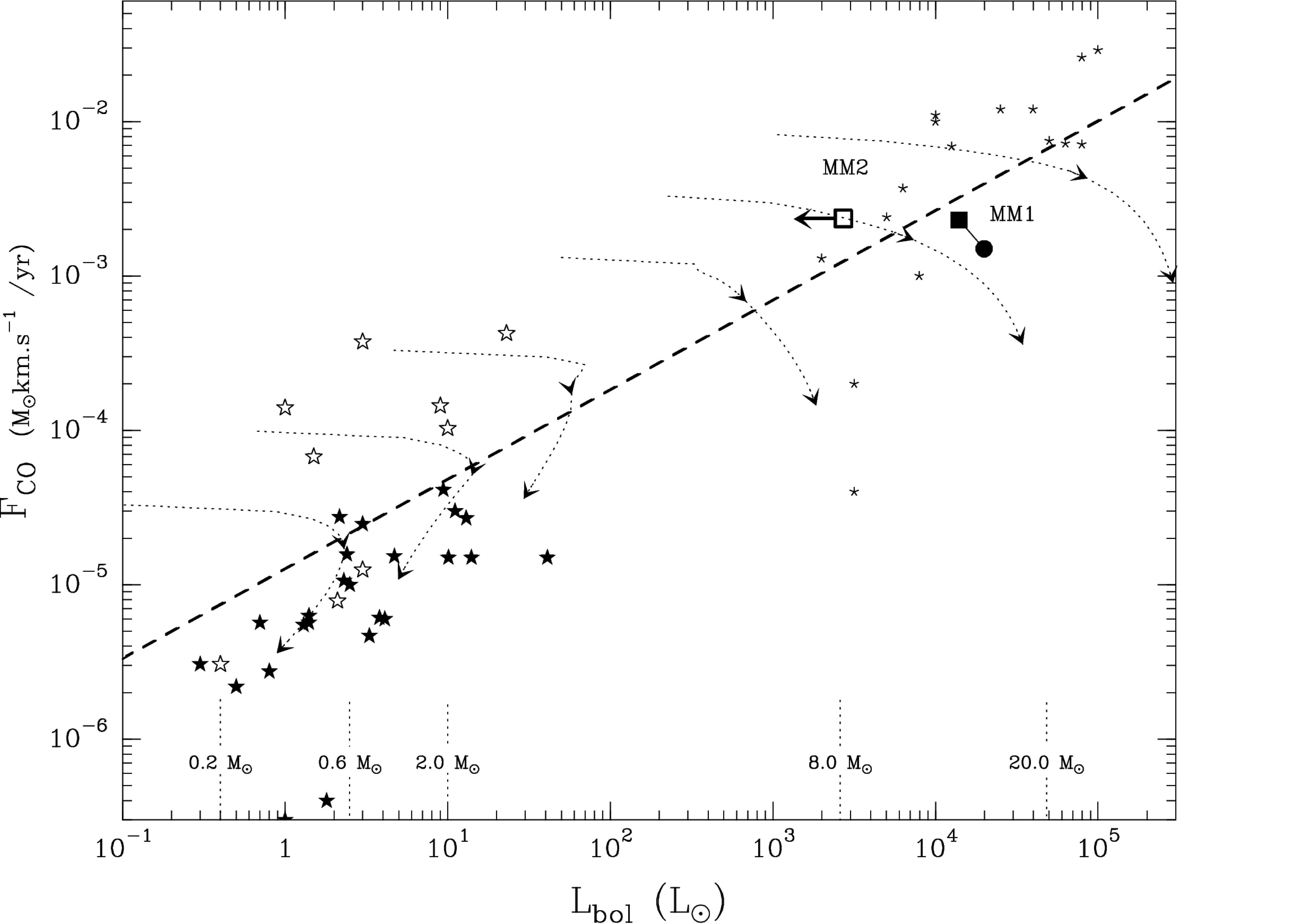}
   \caption{Diagram of the bolometric luminosities versus CO outflow momentum fluxes for protostars adapted from Fig.~5 in \citet{bontemps1996}. 
   Big starred markers (black and
   white): low-mass protostars \citep{bontemps1996}; small starred markers: unresolved massive protostars in HMPOs from \citet{beuther2002b}. The dotted tracks arrows are from
   the \citet{bontemps1996} toy model for masses from 0.2 to 50~\msol. The black circle for MM1 indicates the values obtained by \citet{beuther2002b} for this source, 
   the black squares the values we derived and given in Table~\ref{tab:outflow} and \ref{tab:summary}. The white square indicates the location of the MM2 high-mass protostar. The total luminosity 
   of the MM2 core is taken as an upper limit of the luminosity of the unresolved high-mass protostar.}
   \label{fig:flowevolution}
\end{figure}

One can also use the CO outflow energetics to further discuss the evolutionary
stages of MM1 and MM2. Outflows are believed to be the best indirect tracers of
protostellar accretion. Interestingly enough, while MM2 is not a bright IR
source and is less luminous than MM1, its CO outflow is as powerful as the one
of MM1. This is like Class~0 low-mass protostars which have on average more
powerful outflows than more evolved, and IR bright Class~I objects. It has been
interpreted by \citet{bontemps1996} as due to a decrease of the accretion rate
with time. Extrapolating these results to higher masses, we investigate in
Fig.~\ref{fig:flowevolution} how MM1 and MM2 place in the evolutionary diagram
based on the energetics of CO outflows.  The location of the low-mass protostars
(Class~0 and Class~I) is displayed with stars and is well reproduced by a toy
model (dotted tracks) based on an exponential decrease of the accretion rate
with time \citep[see details in][]{bontemps1996} for stars between 0.2 and
2~\msol. Three additional tracks are given for $M=8$, 20, 50~\msol. One can
note, for instance, that while the low-mass protostars reach a maximum
luminosity which is higher than their final ZAMS luminosity (by a factor of 5
for 0.2~\msol stars) the luminosities of massive protostars are always
increasing.  In this scenario, the dashed line would represent the location of
the transition between Class~0 and Class~I protostars, i.e. the location where
half of the final mass is accreted (first black arrow symbols, second arrow
symbols are for 90~\% of the mass accreted). If this can be extrapolated to
high-mass protostars it indicates that MM2 is younger than MM1 and could host a
high-mass Class~0 like protostar. Note that the HMPO (star symbols) including
MM1 and MM2 are not well resolved as individual collapsing objects. This is
particularly important for MM2 for which the measured luminosity depends on its
continuum fluxes in the millimeter range and therefore directly scales with the
size of the region. The luminosity of MM2 should therefore be seen as an upper
limit of the actual luminosity of the protostar.  In this diagram, MM2 and MM1
are located in the area of the $M=20$~\msol track with MM2 having less than
50~\% of its final mass, and MM1 having of the order of 50~\% of its final mass
accreted.  While the stellar embryo of MM2 is therefore certainly not yet a
ionising star, MM1 could be hosting a weakly ionizing stellar embryo of
$\sim$10~\msol. This could explain why MM1 would appear as the brightest core in
Cygnus X (in Fig.~7 of \citealt{motte2007}) which is not yet associated with a
UC-H{\sc{ii}} region.
	
The overall properties of MM1 and MM2 therefore provide us with a coherent
picture of MM1 and MM2 hosting at least one high-mass protostar which are not
yet creating an UC-H{\sc{ii}} region. MM2 hosts a certainly younger protostar,
and probably slightly less massive, than MM1. The high-mass protostar hosted by
MM1 is a good candidate to be just at the limit of switching on a new
UC-H{\sc{ii}} region. In this scheme, well-known protostars such as AFGL~2591
would be slightly more evolved with an already formed UC-H{\sc{ii}} region and a
much brighter flux in the mid-IR.

	
\subsection{Density and temperature profiles from dust continuum}
	\label{ssec:disc:dens-prof}

\begin{figure}[t!]
  	\centering
  	\includegraphics[width=\columnwidth]{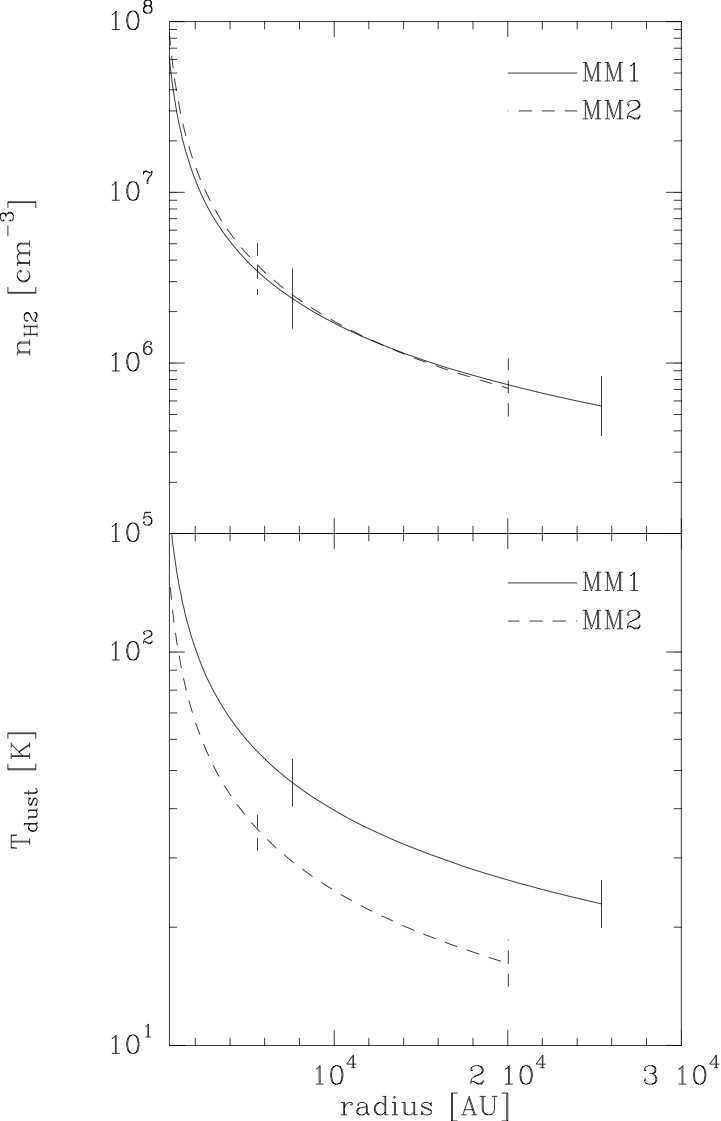}\\
      	\caption{Density (top) and temperature (bottom) profiles used in 1D for molecular line emission modeling of MM1 (plain) and MM2 (dashed). Intervals (vertical lines) represent where 90~\% of the mass is present in the models hence what physical conditions are dominating line emission process of molecules observed.}
         \label{fig:profiles}
   \end{figure}

	From the detailed 1D and 2D analysis of the continuum emission and of the optically thin molecular line modeling, we arrived at the conclusion that a 1D spherical
	geometry is enough to reproduce the molecular line emission. It is only to account for the mid-infrared emission that a 2D geometry is required.
	Our careful analysis showed that the 1D and 2D models were indistinguishable when it comes to model the molecular lines. This is actually not surprising
	since only a very small fraction of the total mass in the core is affected by the 2D inner flattening of the density distribution. 
	Only 11~\msol of the 660~\msol of MM1 is found to be at $T > 100$~K. The bulk
	of mass is at much lower temperatures and does not contribute to the infrared emission (see Fig.~\ref{fig:profiles}).
	
	Facing the same difficulty to reproduce the mid-infrared emission of massive protostars, \citet{vandertak1999} adopted another strategy by using a 1D spherical 
	model with a large inner cavity which could let escape the infrared photons. Our results also validate this approach since it was a way to neglect the 
	inner dust contribution for the molecular line modeling. On the other hand, this approach tends to converge toward large sizes of the cores 
	which are not observed. We believe it is more realistic to impose the sizes of the cores as observed in the dust continuum.
	
	Finally, we note that the use of a self-consistent dust radiative transfer code such as MC3D is ideal to reduce the number of free parameters. On the
	other hand, it imposes some stringent hypotheses on the dust properties by requiring a full opacity curve for dust grains, 
	from infrared to millimeter wavelengths. For instance, we use the MRN distribution of grains which is known not to take into account very well the real
	dust opacity in the millimeter range when ice mantles are present for cold and dense cores \citep{ossenkopf1994}. For SED modeling, this is not very important since dust emission in the millimeter range is optically thin. But it is 
	however important for mass derivation, and we therefore expect that the masses we derived are overestimated due to freeze out onto grains. As a consequence the abundances could be
	underestimated accordingly. The resulting absolute uncertainty can be evaluated by directly deriving total masses
	from the 1.25~mm continuum fluxes of MM1 and MM2 in Table~\ref{table2} using the mass determination by \citet{motte2007} and the average
	temperatures obtained in Table~\ref{tab:summary}. We get 200 and 114~\msol for MM1 and MM2 respectively which are $\sim$3.5 times smaller
	than with MC3D.
	This systematic effect is however a general concern and our resulting abundances are still directly comparable with results from most
	previous similar studies which used the same assumptions. 
		
\subsection{Average abundances from molecular observations.}
	
	Modeling of MM1 and MM2 allowed us to derive absolute molecular abundances. The CS abundances we find, $X \simeq 0.5$-$1.0$~$\times$~10$^{-9}$, are typical values obtained for this type of
	source \citep{pirigov2007}. We note the same result for N$_2$H$^+$ where $X \simeq 3.5$-$6.3$~$\times$~10$^{-10}$ \citep{pirogov2003}, HCO$^+$ with $X \simeq 2.3$-$5.1$~$\times$~10$^{-9}$ \citep{vandertak2000b}, and finally $X$(H$_2$CO) is in the large typical abundance range observed \citep{keane2001,bisschop2007}. All these results fit
	abundance ranges derived from standard chemical evolution modeling as made by \citet{doty2002}: $X$(CS) varying from 1\ttp{-10} to 1\ttp{-8}, $X$(\ndhp) from 1\ttp{-12} to 3\ttp{-10},
	$X$(\hcop) from 1\ttp{-11} to 6\ttp{-9} and $X$(H$_2$CO) from 1\ttp{-10} to 2\ttp{-8}. Thus we conclude that our observations and models do not reveal any abundance anomaly compared to other massive
	protostellar objects and to chemical predictions.
	
\subsection{Depleted CS abundance in the inner regions.}
	\label{ssec:disc:cs}
		
	The modeling of multiple line emission from a single molecular species
        enables us to probe physical conditions inside a source. In our model,
        where no shocks are inserted, the main factor is thermal
        excitation. Modeling of CS transitions ($E_{\textrm{up}}/k$ is equal to
        respectively 7.0, 14.1 and 35.3~K for $J$=2$-$1, $J$=3$-$2 and
        $J$=5$-$4) reveals that for both MM1 and MM2 sources the two lowest CS
        transitions are almost reproduced, whereas the highest one is stronger
        than observed (see Fig.~\ref{fig:csfit-mm1} and
        Fig.~\ref{fig:csfit-mm2}).  A lower abundance of CS in the inner regions
        could explain this discrepancy. Since the gas is cold and dense in
        basically the whole MM1 and MM2 cores a significant depletion of CS onto
        the grains is expected \citep[see][and references
        therein]{tafalla2002}. For the line modeling of the low-mass Class~0
        protostar IRAM~04191, \citet{belloche2002} had also to consider a
        significant CS depletion (by a factor of about 20) to reproduce the
        observed lines. The CS depletion is expected to be still significant
        even for high-mass cores such as MM1 and MM2 because they have not yet
        formed a large hot core region (only a very small fraction of the total
        mass of MM1 is expected to be at $T > 100$~K; see above).  We conclude
        that CS depletion in the inner regions of MM1 and MM2 could be
        responsible for the observed too weak CS(5$-$4) emission which is not
        reproduced by our modeling.
	
\subsection{H$_2$CO: a probe of physical and chemical interactions ?}

Our study clearly shows that H$_2$CO emission of we observed can be modeled with
no radial variation of abundance. The test of an overabundance driven by a hot
core ($T >100$~K) does not show any significant improvement in line fitting. We
can understand it from the low contribution in mass of the hot parts of MM1 and
MM2 (resp. 11~\msol and 3~\msol; see Fig.~\ref{fig:profiles}). Thus we conclude
that H$_2$CO line emission does not show any hot core contribution.
	
However the line emission from MM2 leads to a greater abundance of H$_2$CO than
MM1, with a need for a hotter inner part although it is the colder and the
younger of the two sources.  The line emission from MM2 could be affected by a
significant contribution of outflow shocks in the inner envelope of the
protostar. This idea is supported by the higher turbulent velocity observed in
MM2 ($\varv_T = 1.7$~\kms\ in contrast to 1.0~\kms\ for MM1) and its internal
para-to-ortho ratio greater than 1, suggesting recent chemical activity
\citep{kahane1984} driven in these shocks.

\subsection{Chemical evolution of high-mass cores.}

We finally wish to speculate that the chemical differences between MM1 and MM2
are related to their different evolutionary stages.  The derived abundances,
obtained with the same modeling process and radiative transfer code, show that
$X$(CS) is 2 times higher for MM1 than for MM2, and 5 times lower than in the
probably more evolved AFGL~2591 high-mass protostar, suggesting that the CS
abundance might increase with time. Interestingly enough, the
\citet{johnstone2003} study of sub-millimeter sources in Orion was suggesting
the same trend. The chemical models by \citet{wakelam2004} actually predict such
an evolution of the CS abundance in protostars: a constant production of CS is
expected at low temperatures (T$<$100~K) thus increasing the abundance with the
source evolution. Moreover, when the hot core region develops, the release of CS
from the grains may even increase more the CS abundance. This could be the case
for the high abundances in AFGL2591 and in the sources observed by
\citet{hatchell2003}.
          
In contrast, $X$(N$_2$H$^+$) is found to decrease with time; the N$_2$H$^+$
abundance inside AFGL~2591 shows a clear trend of decrease from MM2 to
AFGL~2591: $X$(N$_2$H$^+$)$_{\mathrm{AFGL~2591}}$ =
0.1~$\times$~$X$(N$_2$H$^+$)$_{\mathrm{MM1}}$ =
0.05~$\times$~$X$(N$_2$H$^+$)$_{\mathrm{MM2}}$. This molecule hardly depletes on
cold dust grains and is rapidly destroyed at warm temperatures. This behaviour
is not unique and has been already observed in high-mass protostellar objects
\citep{reid2007}.  We suggest that the CS and N$_2$H$^+$, {\it i.e.} abundance ratio
may be a good tracer of protostellar evolution, but more observations and
modeling are required to test this hypothesis. 

\section{Conclusions}

Here we summarize our conclusions on the massive protostellar objects MM1 and
MM2 of the IRAS~18151$-$1208 region.

\begin{enumerate}
\item The three cores of the region, MM1, MM2 and MM3 are physically linked and
  have probably been formed from a single parental cloud or clump.
\item We detected for the first time a CO outflow driven by MM2. It clearly
  establishes the protostellar nature of MM2. In contrast MM3 does not show any
  outflow activity and is therefore most probably a pre-stellar core.
\item Following the evolutionary scheme discussed in \citet{motte2007}, MM1 is a
  high luminosity IR (or IR-bright) massive protostar while MM2 is an IR-quiet
  massive protostar.
\item We have established that while an inner flattening of the matter
  distribution is required to reproduce the SED of MM1, a simple 1D spherical
  geometry is enough to well model molecular line observations. In contrast, MM2
  does not even require a inner flattening since it is not detected in the
  mid-infrared range.
\item A significant depletion of CS in the inner parts of the MM1 and MM2 cores
  is required to fully reproduce the observed CS line emission.
\item We find that the abundance ratio between CS and N$_2$H$^+$ could be a very
  good evolution tracer for high-mass protostellar cores hosting high-mass
  protostars.
\end{enumerate}

\begin{acknowledgements}
  We would like to thank Henrik~Beuther for his useful and precise comments as
  referee and for kindly providing his 1.2~mm continuum map data of the region
  that we have used for the figure of the whole region
  (Fig.~\ref{fig:msx-map}). They both greatly improved the general content of
  this paper.  We also thank Jonathan Braine and Nicola Schneider for their
  corrections.
\end{acknowledgements}


\bibliography{biblio}
\bibliographystyle{aa}

\end{document}